\documentclass[a4paper,11pt]{article}
\usepackage[utf8]{inputenc} 
\usepackage[T1]{fontenc} 
\usepackage{graphicx} 
\usepackage{amsmath} 
\usepackage{geometry} 
\usepackage{setspace} 
\usepackage[style=apa, backend=biber]{biblatex} 
\usepackage{longtable} 
\usepackage{booktabs}
\usepackage{fancyhdr}
\usepackage{array}
\usepackage{hyperref}
\usepackage{soul}
\usepackage{float}
\usepackage{caption}
\usepackage{subcaption}
\usepackage{etoolbox}
\usepackage{placeins}
\usepackage{adjustbox}
\usepackage{float}

\title{\textbf{Disentangling the Distributional Effects of Financial Shocks in the Euro Area}\thanks{We thank Dario Bonciani, Cristiano Cantore, Juan José Dolado, Luca Fanelli, Francesco Furlanetto, Carlo Galli, Francesco Lucidi, Lorenzo Mori, Salvatore Nisticò, Evi Pappa, Valeria Patella, Michele Raitano, Luca Riva, Emircan Yurdagul, and all participants at the 5th Sailing the Macro Workshop, the 9th International PhD Meeting in Economics, the Macro-Metrics Sapienza Reading Group, the Macro-Working Progress Carlos III Reading Group, and the Manic Mondays Seminars for helpful comments.\\This paper is based on data from Eurostat, EU-SILC, 2024. The  responsibility for all conclusions drawn from the data lies entirely with the authors. This paper uses data from the Eurosystem Household Finance and Consumption Survey.\\We acknowledge financial support from the Sapienza University of Rome under the Avvio alla Ricerca 2024 funding program.}}

\author{
    Miloš Ciganović\textsuperscript{\thanks{Department of Economics and Law, Sapienza University of Rome. Emails: \\milos.ciganovic@uniroma1.it\\elena.scolagagliardi@uniroma1.it\\massimiliano.tancioni@uniroma1.it}} \and 
    Elena Scola Gagliardi\textsuperscript{\thefootnote} \and 
    Massimiliano Tancioni\textsuperscript{\thefootnote}
}

\date{October, 2025}

\begin{document}

\maketitle
\begin{abstract}

We estimate the dynamic distributional effects of financial shocks in the Euro Area using survey-based microdata on personal incomes. We find that positive financial shocks increase inequality, with heterogeneity across different income groups. Much of the response emerges in the tails of the income distribution. By decomposing individual incomes into financial and labor components, we identify two distinct transmission mechanisms: financial income inequality rises, likely due to differences in asset holdings. In contrast, labor income inequality increases through a skill premium channel. We then consider a nonlinear model framework, distinguishing the sign of the shock, allowing us to document the presence of asymmetric effects. While positive shocks lead to income disparities, adverse shocks have the opposite effect. Notably, middle-income groups are only affected following a negative shock, highlighting differential vulnerabilities across the income distribution.

\end{abstract}



\FloatBarrier 

\newpage

\section{Introduction}

\begin{figure}
    \centering
    \includegraphics[width=0.7\linewidth]{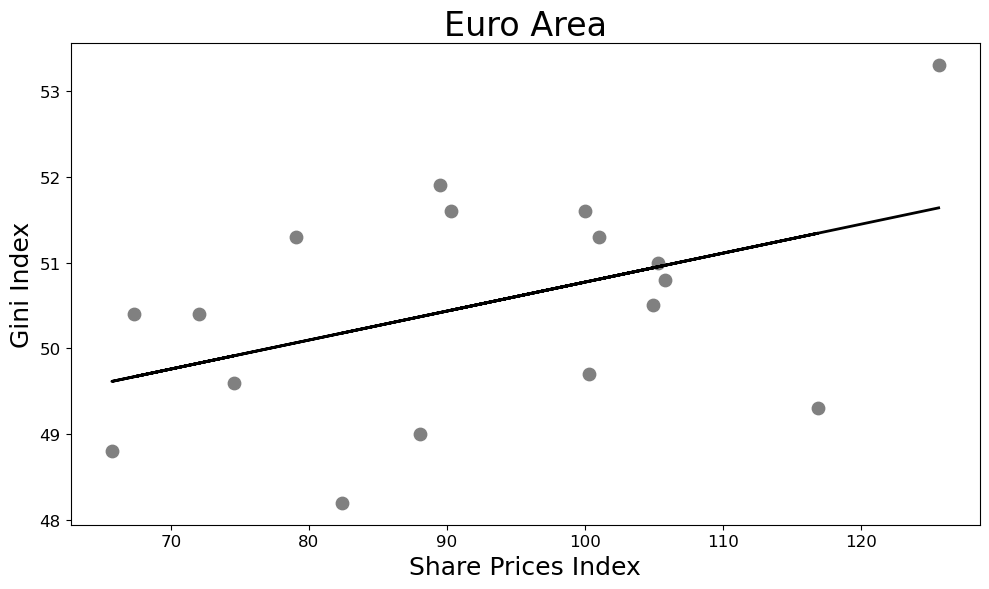}
    \caption{Share Prices and Income Inequality in the Euro Area: Unconditional Evidence}
 
    \footnotesize
    The figure shows the scatter plot of income inequality and share prices in the Eurozone (grey dots). The fitted linear trend (black solid line) indicates a positive association.
    \label{fig:scatter_plot}

\end{figure}

What are the effects of financial shocks on income inequality? What are their transmission mechanisms? Due to the recent dynamics of inequality observed in developed economies and its intricate links with the business cycle  \parencite{stiglitz2015macroeconomic}, the topic has garnered growing interest among researchers and policymakers \parencite{dabla2015causes}.\\ The inequality literature addresses financial factors as key contributors to these developments (i.a. \cite{demirgucc2009finance}). Existing studies in this field have explored the relationship between inequality and specific financial dimensions, yielding mixed results \parencite{de2017finance}. The specific domains addressed by this literature are assimilable under three perspectives:  the role of financial development (i.a. \cite{kappel2010effects, zhang2019financial}), of financial liberalization (i.a. \cite{ furceri2015capital}),\footnote{As noted by \textcite{ABIAD2008270}, financial liberalization and financial development are conceptually distinct. The former refers to a reduction in government intervention and a greater role for market forces. In contrast, the latter captures the expansion of financial activity \parencite{de2017finance}.} and of banking crises (i.a. \cite{atkinson2011economic}).\\ Turning to the macroeconomic literature, the interest in the role of financial shocks has focused mainly on standard macroeconomic endpoints \parencite{gilchrist2009credit, peersman2011bank, meeks2012credit, peersman2014shocks, caldara2016macroeconomic, gambetti2017loan, furlanetto2019identification, brianti2025financial}, leaving their role in affecting the distributional dynamics nearly unexplored. Recent contributions in the macro literature have examined how financial frictions can amplify the distributional consequences of monetary policy shocks \parencite{caldara2019monetary, ferlaino2024does}.\\ 
Our paper contributes to filling the apparent gap between the inequality and macro literature by examining the effects of financial shocks on income inequality and their transmission mechanics in the Euro Area. To the best of our knowledge, no study has yet isolated the direct impact of unexpected variations in stock prices on measures of inequality. Moreover, the focus on the European macroeconomy, which remains relatively under-explored, can offer new and potentially challenging results for the consolidated macro literature \parencite{altavilla2024research}.\\ Figure \ref{fig:scatter_plot} presents unconditional evidence of a positive correlation between share prices and income inequality in the Eurozone. Although suggestive at best, this association motivates our focus on the implications for inequality of idiosyncratic stock market movements.

The existing empirical literature suggests that transitioning from unconditional correlations to a conditional analysis can be challenging when the shock of interest originates in the financial sector. Even if there is little dispute about the fact that observed movements in stock prices explain a relevant fraction of macroeconomic fluctuations, financial shocks are indistinguishable from uncertainty shocks with standard identification methods (\cite{bloom2009impact, caldara2016macroeconomic, furlanetto2019identification, ludvigson2021uncertainty}). Their simultaneity precludes the recourse to contemporaneous exclusion restrictions, and - absent specific financial measures for the Euro Area countries - credible set (sign)-based restrictions are hardly conceivable for shocks that tend to affect the economy in the same direction (\cite{caldara2016macroeconomic, ludvigson2021uncertainty, brianti2025financial}). Moreover, identification methods based on quantitative restrictions (\cite{caldara2016macroeconomic, carriero2025max}) might fail because the magnitude of the stock price responses or its conditional variances are not necessarily maximized by the own (financial) shock (\cite{brianti2025financial}).\\
We address these empirical identification issues by adopting a simple two-step modelling approach, which involves controlling for uncertainty in the second step only. In the first step, we recover the structural shocks using a panel vector autoregressive methodology (P-SVAR) identified with sign restrictions. In the second step, we include the (possibly confounded) financial shock in a set of \textcite{jorda2005estimation}'s panel local projections (P-LPs) to recover the impulse responses of interest.\footnote{Such a two-step strategy resembles that recently adopted by \textcite{forni2024nonlinear} and \textcite{lucidi2024effects}.} The P-SVAR is identified with the baseline sign restrictions proposed in \textcite{furlanetto2019identification}, distinguishing the "uncertainty-confounded" financial shocks from the other sources of macroeconomic variability. We thus include contemporaneous and lagged measures of financial and macroeconomic uncertainty available at the EU country level in the control set of the second step P-LPs.\footnote{We consider two measures of uncertainty and financial market volatility: the smoothed World Uncertainty Index developed by \textcite{ahir2022world}, and the Country-Level Index of Financial Stress (CLIFS) constructed by the European Central Bank (ECB).}\\
The baseline results indicate that financial shocks increase income inequality in the Euro Area. This finding contrasts with the evidence for the United States on the cyclicality of inequality \parencite{guvenen2014nature}. However, it aligns with recent evidence of procyclical inequality in Europe — Norway \parencite{Bergholt2024A} and Germany \parencite{Ettmeier2024Functional} — specifying that financial shocks primarily drive such pro-cyclicality. This result highlights the importance of distinguishing between different sources of business cycle fluctuations when examining their role in the distributional dynamics.\\ 
To better understand the conditional distributional dynamics, we then focus the analysis on specific segments of the income distribution.\footnote{Recent macroeconomic literature has increasingly emphasized the role of agent heterogeneity and distributional considerations, primarily due to their policy relevance. A prominent example is the growing body of research on Heterogeneous Agent New Keynesian (HANK) models, which explicitly account for household heterogeneity in studying macroeconomic dynamics. See, for instance, \textcite{kaplan2018microeconomic}. Understanding these dynamics is essential, as macroeconomic shocks may have disparate effects across specific segments of the income distribution, necessitating targeted government interventions and prompting public support for economic policy.} We utilize survey data from the European Union Statistics on Income and Living Conditions (EU-SILC) to examine the conditional dynamics of interquintile Gini indices, assessing how the income distribution is affected by financial shocks. The quintile-specific results show that the effects of financial shocks are more pronounced at the tails of the income distribution.\\
To shed further light on the transmission mechanisms, we decompose individual income into labor and financial components, allowing us to track how inequality in each responds to financial shocks. While this distinction has attracted growing attention in the study of aggregate macroeconomic shocks — particularly in the monetary policy literature (see, e.g., \textcite{amberg2022five, andersen2023monetary}) — it remains largely unexplored in the context of financial shocks. We document that a key channel through which financial shocks influence income inequality is the unequal distribution of financial asset holdings. This result aligns with the financial development literature, which distinguishes between the extensive margin — referring to access to financial markets — and the intensive margin, which captures the depth of participation within them. If a financial shock expands market access (i.e., affects the extensive margin), it may reduce inequality. Conversely, if its benefits accrue primarily to those already integrated into financial markets (i.e., affect the intensive margin), inequality may increase. By focusing on positive financial incomes, we provide evidence consistent with a mechanism operating through the intensive margin.\\ 
A more nuanced question is whether financial shocks also affect labor income inequality, and through which mechanisms. We find that labor income inequality responds positively to financial shocks over medium-term horizons. While multiple (and not mutually exclusive) factors may underlie this relationship, we provide evidence for one specific channel, related to an evident pro-cyclicality of top earners' incomes within the distribution. In particular, positive financial shocks lead to higher remuneration for senior executives and top management, thereby contributing to greater labor income inequality. Furthermore, drawing on the International Labour Organization (ILO) classification of occupational tasks (ISCO-08) available in our dataset, we show that this effect is conceivable through the lens of a skill premium: financial shocks disproportionately increase the wages of high-skilled workers relative to those of low-skilled workers.\\
To align our analysis with the theoretical and empirical macro literature \parencite{brunnermeier2014macroeconomic, barnichon2022effects} on the emergence of asymmetric effects from positive and negative financial shocks, we then emphasize the effects of adverse (negative) financial shocks. From the nonlinear estimates, we show that positive and negative financial shocks affect inequality in opposite directions: while positive shocks increase inequality, negative shocks lead to a gradual and persistent decline. When decomposing inequality using interquintile Gini indices, we find that the middle of the income distribution is also significantly affected following a negative financial shock.\\
These results are robust to several control dimensions. Specifically, we test the dependence of results on the peculiar identifying assumptions, first removing the uncertainty controls from the LPs and then considering a standard recursive identification scheme for the P-SVAR, with the stock price variable ordered last as the fastest-moving variable. We then test the relevance of movements in stock prices originating in the the credit sector (\cite{liu2013econometrica, justiniano2015red}), where credit shocks are separated from financial shocks considering a variant of the sign restrictions strategy in \textcite{furlanetto2019identification}, based on a magnitude restriction. Concerning the inequality endpoint, we consider the response of incomes in the 90th and 95th percentiles as an alternative to the Gini index. Finally, we test robustness to changes in the time and sectional dimensions of the sample. We first exclude the covid-19 period (thus reducing the temporal dimension to 2019:4) and then remove Latvia and the Netherlands from the sample, as they have experienced extreme financial shocks. Results remain qualitatively unchanged across all the control dimensions.\\       
Our paper relates to two separate streams of literature: the extensive literature on inequality, specifically addressing the role of financial factors, and the emerging macroeconomic literature, which examines the impact of macroeconomic shocks on distributional dynamics.\\
From the first perspective, a closely related contribution is \textcite{gocmen2025private}, who link the rise in United States income concentration to high-net-worth individuals earning excess returns on early-stage private investments. While their focus differs, their findings complement ours by showing how individual portfolio choices, driven by economic conditions, can shape income trajectories and distributional outcomes. These results align with our previous findings that financial shocks exacerbate disparities in financial income. More broadly, our work contributes to the literature on financial development and inequality (e.g., \cite{de2017finance}) and echoes the intensive margin mechanism emphasized in \textcite{greenwood1990financial}.\\
From the second perspective, we relate to several recent studies in the macro literature. \textcite{amberg2022five} document how monetary policy shocks affect different income groups in Sweden, with responses shaped by income composition and heterogeneity. Notably, they find that expansionary shocks tend to affect both low- and high-income individuals more than those in the middle of the distribution — a pattern consistent with our findings. Similarly, \textcite{andersen2023monetary} study the effects of monetary policy on income, wealth, and consumption in Denmark, emphasizing the importance of distinguishing income components. These studies, like ours, highlight the importance of distinguishing between labor and financial income to gain a deeper understanding of the transmission of macroeconomic shocks. This reasoning also aligns with \textcite{gornemann2021doves}, who show that the income composition is a key determinant of who gains or loses following aggregate shocks. Finally, \textcite{dolado2021monetary} demonstrate that unexpected monetary easing increases labor income inequality between high- and low-skilled workers in the United States. We view this as complementary to our interpretation of labor income responses to financial shocks through the lens of the skill premium.\\
The remainder of the paper proceeds as follows. Sections \ref{data} and \ref{sec:empirical_strategy} outline the data and the empirical strategy adopted in the analyses. Section \ref{baseline_evidence} presents the main results. Section \ref{sign_effect} focuses on the asymmetric inequality implications of financial shocks, and Section \ref{conclusions} concludes.

\section{Data} \label{data}

We consider quarterly data for a panel of Euro Area countries over the period 2006:1–2023:4.\footnote{Countries included in our sample are: Austria (AT), Belgium (BE), Germany (DE), Estonia (EE), Greece (EL), Spain (ES), Finland (FI), France (FR), Italy (IT), Lithuania (LT), Luxembourg (LU), Latvia (LV), Netherlands (NL), Portugal (PT), Slovenia (SI), and Slovakia (SK). } The analysis exploits the cross-sectional and time series information contained in two types of data: (i) Microdata for income-related variables, and (ii) macroeconomic time series.
 
\subsection{Microdata-based income and inequality measures}

To construct our inequality measures, we use microdata from the EU-SILC survey, which provides harmonized cross-sectional and longitudinal data on income, poverty, and living conditions across EU member states.\\
The analysis considers both an aggregate measure of income inequality in the Euro Area and inequality measures across specific segments of the income distribution (quintiles). Furthermore, we distinguish between financial and labor income inequality and specific segments of the ILO's classification of tasks, the latter to identify the skill specificities in the transmission of financial shocks to labor income inequality.\\
We construct our income variable based on the definition of market income provided in the Methodological Guidelines and Description of EU-SILC Target Variables issued by the European Commission (2023 version). Accordingly, to better capture income inequality, we use the equivalised version of income by applying the modified OECD equivalence scale. This procedure involves computing total household income and then dividing it by an equivalence factor that reflects household composition. The equivalence factor assigns a weight of 1.0 to the first adult, 0.5 to the second adult, and to each additional person aged 14 or over, and 0.3 to a child under the age of 14. Based on this definition of individual income, we compute Gini indices by country, as well as the other inequality measures. We then distinguish between financial and labor income. To capture the intensive margin, we consider only positive values of income.\\
To investigate the role of the skill premium, we construct a skill-based income measure defined as the ratio of the average annual income of high-skilled workers to that of low-skilled workers. Skill levels follow the ILO definition, based on the ISCO classification of occupations. The reference version is ISCO-08. This framework assigns skill levels ranging from 1 (lowest) to 4 (highest). We define high-skilled workers as those in levels 3 or 4, which correspond to the following occupational groups: Managers, Legislators, Professionals, and Technicians and Associate Professionals. We consider only individuals of working age (15–64 years) in this section.

 \subsection{Macroeconomic aggregates}

In addition to income-related variables, we use six macroeconomic time series in the baseline model estimates: GDP, GDP deflator, investment-to-output ratio, stock prices, interest rates, and financial deepening. Data are primarily sourced from the Eurostat Database, further described in the Data Section (Appendix \ref{data_app}).

\section{Empirical strategy}
\label{sec:empirical_strategy}

To provide empirical evidence on the effects of financial shocks on income inequality, we employ a two-step methodology \parencite{forni2024nonlinear, lucidi2024effects}. The first step involves recovering the country-specific structural shocks of interest from a (pooled) P-SVAR. In contrast, the second focuses on estimating the impulse responses using a panel version of the LP method, as described by \textcite{jorda2005estimation}. In the first step, we identify the financial shock, along with all other shocks, following the baseline strategy adopted in \textcite{furlanetto2019identification}. In the second step, P-LPs, the confounding effects stemming from changes in uncertainty are controlled for by including macroeconomic and uncertainty measures in the set of controls, along with section- and time-specific effects.\\
The first reason for adopting the two-step method is the impossibility of separating the financial shock from uncertainty with set-based restrictions in the absence of country-level data on the external finance premium. A second reason is the simplicity of the LP methodology: LPs rely on a single regression estimation where the dependent variable is the specific inequality measure at different horizons, thereby minimizing the number of parameters to be estimated. A third reason for our approach stems from considerations about the variance-bias trade-off characterizing the choice between LPs and VARs, with bias being the primary concern when dealing with structural estimates \parencite{li2024local}. Consequently, coverage probability issues related to confidence intervals construction are avoided \parencite{olea2024double}. A more formal description of our baseline two-step strategy is given below.

\subsection{Identification} \label{sec: step_one}

Recent literature indicates that movements in stock prices are often accompanied by significant changes in uncertainty measures \parencite{bloom2009impact}. As long as financial and uncertainty shocks tend to affect the macroeconomy and financial indicators simultaneously and in the same direction, it is unclear whether standard identification methods based on exclusion and/or set (sign) restrictions can be defended on theoretical or statistical grounds (\cite{caldara2016macroeconomic, ludvigson2021uncertainty, brianti2025financial}).\\ In the context of set-based restrictions, financial shocks and uncertainty shocks are separable by assuming that the response of the external finance premium relative to uncertainty is higher for uncertainty shocks with respect to adverse financial shocks (\cite{ShinZhong2018, furlanetto2019identification}). Unfortunately, such a strategy is not viable in our setting, as an off-the-shelf measure of the excess bond premium is not available for the European countries considered in the analysis, unlike uncertainty measures.\\ Alternative identification methods based on quantitative restrictions, as max share penalties (\cite{caldara2016macroeconomic}; \cite{carriero2025max}), magnitude inequality constraints (\cite{ludvigson2017nber}), or heteroskedasticity-based restrictions (\cite{lutkephol2021ectj}), cannot be considered safe alternative strategies, since stock price responses or variance magnitudes are not necessarily larger for idiosyncratic sources of variability (\cite{brianti2025financial}).\\
For these reasons, we address the empirical identification issue with the simple two-step modelling approach sketched above, which involves controlling for country-specific uncertainty in the second step only.\\
We borrow the identification strategy of financial shocks from the work of \textcite{furlanetto2019identification}, that provides a general characterization of financial shocks based on sign restrictions on the impact effects matrix \parencite{faust1998robustness, canova2002monetary, peersman2005caused, uhlig2005effects, fry2011sign} consistent with most financial innovations studied in the macroeconomic literature: a positive financial shock is a demand shock leading to an increase in all variables in the PVAR. Such a shock differs from a general demand shock in that it assumes a pro-cyclical investment-to-output ratio response, and from an investment-specific shock in that the latter generates a negative response of the stock price index. The opposite sign of the responses of the macro variables to interest rate changes separates the monetary policy shock from the financial shock (Table \ref{tab:restrictions_id1})\footnote{We present the identification scheme for clarity purposes only. A more detailed exposition of the set-based identification scheme can be found in the source paper. The results obtained from the P-SVAR are available upon request from the authors. Details on the model specification are provided in Appendix \ref{bayesian_spec}.}.

\begin{table}[h!]
\centering
\caption{Identification Scheme (\textcite{furlanetto2019identification})}
\begin{tabular}{lccccc}
\toprule
\textbf{}       & \textbf{Supply} & \textbf{Demand} & \textbf{Monetary} & \textbf{Investment} & \textbf{Financial} \\
\midrule
GDP             & +               & +               & +                 & +                   & +                   \\
Prices          & $-$             & +               & +                 & +                   & +                   \\
Interest rate   &               & +               & $-$               & +                 & +                   \\
Investment/output &             & $-$             &                 & +                   & +                   \\
Stock prices    & +               &               &                 & $-$                 & +                   \\
\bottomrule
\end{tabular}
\vspace{0.5cm}
\begin{minipage}{0.9\textwidth}
\small
\end{minipage}\\
\footnotesize 
    The table describes set-based restrictions imposed at impact for different variables (rows) in response to the shocks (columns). A blank space indicates an unrestricted response.
\label{tab:restrictions_id1}
\end{table}

\subsection{P-LP-based impulse responses} \label{sec: second_step}

Once the possibly "contaminated" financial shocks are recovered from the P-SVAR, impulse responses are directly estimated through lag-augmented P-LPs \parencite{plagborg2021local}, including the contemporaneous and lagged values of the macroeconomic and financial uncertainty measures as controls. The P-LPs also consider controls for section and time-invariant confounding factors. Specifically, we adopt the Two-Way Mundlak (TWM) approach \parencite{wooldridge2021two} to the Two-Way Fixed Effects estimator.\\
Let \(\varepsilon_{i,t}\) be the country-specific financial shock series, identified within the P-SVAR. The \(h=0,1,...,H\) periods ahead impulse response function for the outcome variables of interest \(y_{i,t}\) (measures of income inequality) is obtained  considering the following P-LP specification: 

\begin{equation}\label{panel_lp}
    y_{i, t+h} = \beta_{h} \varepsilon_{i,t}\ + \sum_{l=1}^{p}{\gamma_{h,l} X_{i,t-l}} + \sum_{l=0}^{p}{\psi_{h,l} U_{i,t-l}} + \vartheta \bar{x}_i + \rho \bar{x}_t + v_{i,t+h},
\end{equation}
where \(\beta_h\) are the impulse response parameters of interest, \(X_{i,t-l}\) is a vector of controls including lags of the outcome and shock variable together with the lags of all the variables included in the P-SVAR, and \(U_{i,t-l}\) contains the contemporaneous and lagged uncertainty controls.\footnote{We fix the order of lag \(p\) to four. We have verified that results are robust to changes in the lag order.} Given that the time-evolving degree of financialization may act as a relevant source of cross-country heterogeneity, we also include in the controls set \(X_{i,t-l}\) a measure of financial deepening, defined as the ratio of total financial sector assets to GDP.\footnote{Previous studies assessing the effects of financial deepening have typically relied on narrower indicators, such as the ratio of money to GDP \parencite{li1998explaining}, or quasi-money to GDP \parencite{milanovic2005can}. However, to better capture the overall size of the financial system, we adopt a broader indicator that includes a wider range of financial instruments.} The variables \(\bar{x}_i\) and \(\bar{x}_t\) denote the TWM unit and time effects.\\ With this starting specification, the P-LP closely resembles the equation of interest in the P-SVAR, ensuring the consistency of the P-LP with the P-SVAR \parencite{montiel2021local, olea2024double}.\\
The equivalent within transformation of the P-LP is the following:

\begin{equation}
    \ddot{y}_{i,t+h} = \beta_{h} \ddot{\varepsilon}_{i,t} + \sum_{l=1}^{p}{\gamma_{h,l}} \ddot{X}_{i,t-l} + \sum_{l=0}^{p}{\psi_{h,l}} \ddot{U}_{i,t-l} +
    v_{i,t+h},
\end{equation}
where double-dotted letters represent double demeaned variables. More specifically, the generic variable \(\ddot{x}\) is defined as \(\ddot{x}_{i,t} = x_{i,t} + \bar{x}_{i,t} - \bar{x}_i - \bar{x}_t\), where \(x_{i,t}\) is the original variable, and \(\bar{x}_{i,t}\) is its overall mean. Since the pooled P-SVAR already controls for the sectional variability through section-specific constant terms, the structural shocks (\(\ddot{\varepsilon}_{i,t}\)) enter only in a time-demeaned form.\\ %Given the equivalence established by the Frisch-Waugh-Lovell theorem, country-specific uncertainty controls are implemented by residualizing the outcome and shock variables through two auxiliary regressions of these variables on the contemporaneous and lagged values of the uncertainty measures.\\
To account for potential correlation among residuals, standard errors are adjusted considering the \textcite{driscoll} method. The maximum autocorrelation window (\(m\)) is set to \(h + 1\), with \(h\) being the horizon of the impulse response function \parencite{jorda2005estimation, tenreyro}.\footnote{For robustness, we consider other maximum autocorrelation window values. Specifically, we consider \(m = p + 1 \), with \(p\) being the number of lags, and \(m = p + h + 1 \).}\\

\section{The effects of financial shocks on income inequality}\label{baseline_evidence}

\begin{figure}[H]
    \centering
    \includegraphics[width=0.8\linewidth]{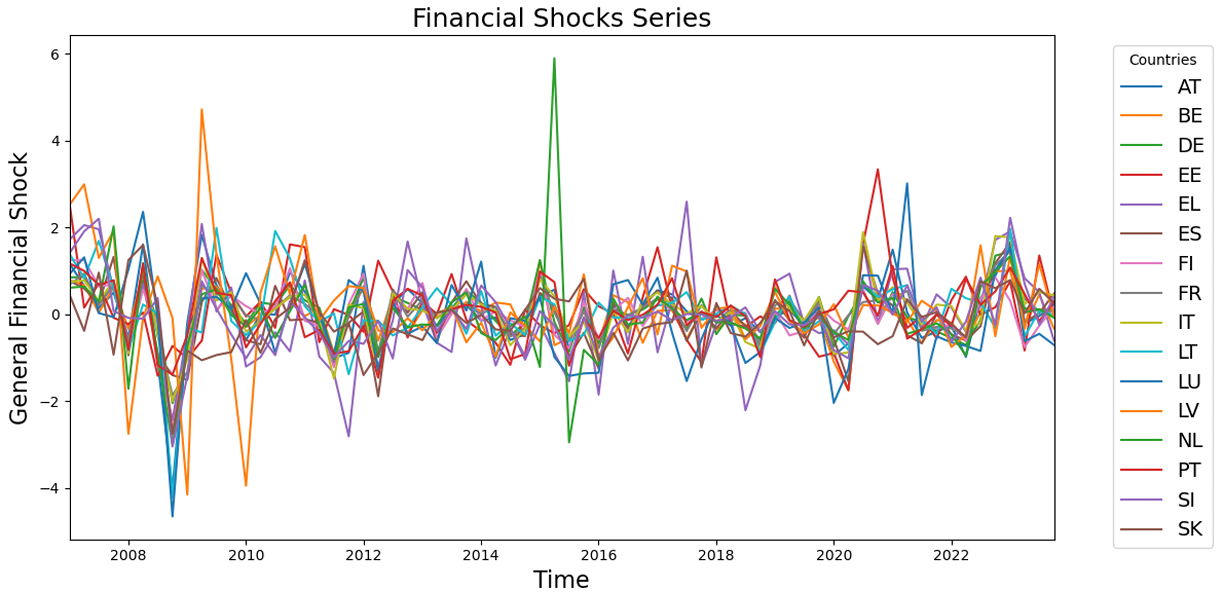} 
    \caption{Financial shocks series} 
    \vspace{0.15cm}
\footnotesize  
Country-level time series of financial shocks in the Euro Area, identified as in Table \ref{tab:restrictions_id1}.
    \label{fig:general_financial_shock}
\end{figure}

This section provides baseline empirical evidence on how financial shocks affect income inequality in the Euro Area. We present results at both the aggregate level and for specific segments of the income distribution. Furthermore, we examine the underlying drivers of the responses by decomposing income into financial and labor components, allowing us to investigate the roles of two key potential transmission mechanisms: financial asset holdings and the skill premium.\\
Figure \ref{fig:general_financial_shock} displays the country-level time series of financial shocks estimated with the P-SVAR in the first step of our empirical strategy. Each line represents one of the Euro Area countries included in the sample. The shocks fluctuate around zero, consistent with their interpretation of unexpected financial disturbances.\\
The country-specific shocks effectively capture the significant episodes of financial market volatility. A clear, common pattern emerges around 2008–2009, coinciding with the Global Financial Crisis, where many countries experienced large adverse shocks. A second wave of higher dispersion occurs during 2020–2021, consistent with the financial market turbulence triggered by the covid-19 pandemic.\\
While most countries exhibit moderate variation, some display more pronounced patterns. In particular, Latvia (orange) and the Netherlands (green) stand out for the magnitude of their shocks over specific periods.\footnote{Results remain qualitatively unchanged when these two countries are excluded from the analysis (Figure \ref{fig:excl_LV_NL}, Appendix \ref{excl_LV_NL}).}\\
Overall, the estimated country-specific shocks validate the identification strategy and reflect meaningful variations in financial conditions at the country level.

\subsection{Financial shocks and income inequality. Baseline results} \label{res_general}

\begin{figure}[htbp]
        \centering
        \includegraphics[width=0.5\linewidth]{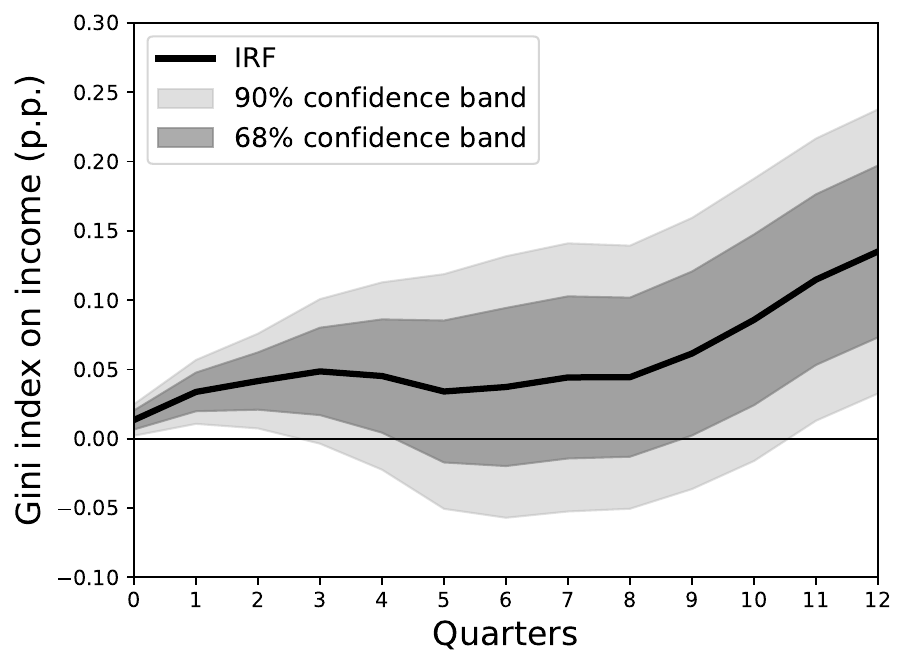}
    \caption{Income inequality responses in the Euro Area}
    \vspace{0.15cm}
\footnotesize 
The figure illustrates the response of income inequality following a one-standard-deviation positive financial shock. Full lines are point estimates; shaded areas indicate 90\% and 68\% confidence bands computed from panel HAC standard errors.
    \label{fig:gini_euro_aggregate}
\end{figure}

Figure \ref{fig:gini_euro_aggregate} displays the response of the income Gini index after a one-standard-deviation positive financial shock occurs. The solid black line shows the estimated impulse response. Shaded areas denote confidence intervals—68\% in dark gray, 90\% in light gray.\\
The Gini index rises on impact and continues to increase over time, peaking at 0.14 percentage points by the end of the third year, indicating that a positive financial shock - associated with rising share prices - leads to an increase in income inequality. This basic finding aligns with the evidence of pro-cyclical inequality in Europe, recently documented for Norway by \textcite{Bergholt2024A} and for Germany by \textcite{Ettmeier2024Functional}. From this perspective, our result confirms the apparent contrast with evidence from the United States, indicating a counter-cyclical behavior of inequality.\\ 
To assess whether our main result reflects the effects of general aggregate demand shocks or is specific to financial shocks, we also examine the responses of inequality to other shocks identified in the S-PVAR. Results show that in the Euro Area, financial shocks are the primary source of the observed pro-cyclicality. In contrast, other demand shocks have neutral or even equalizing effects (Appendix \ref{additional_results}).\footnote{The result for the generic demand shock (second column of Table \ref{tab:restrictions_id1}) is in the Appendix \ref{additional_results}, Figure \ref{fig:inequ_dem_shock}. Due to space constraints, the remaining IRFs are available upon request. We thank Francesco Furlanetto for suggesting this exercise and for highlighting its relevance and connection to the literature. Please note that, since the shock of interest here is not financial, the measures of uncertainty are not considered among the control variables.} This result highlights the importance of the composition of business cycle shocks in shaping the inequality–cycle relationship across countries.

Figures \ref{fig:ineq_no_uncertainity} and \ref{fig:chol} in the Appendix (\ref{ineq_no_uncertainity_sec} and \ref{alternative_id}) show that the main findings hold when removing the uncertainty controls from the P-LP and when using a recursive (Cholesky) identification, where the ordering of the variables follows that of Table \ref{tab:restrictions_id1}. The results of the robustness checks considering alternative measures of the distribution \footnote{The choice of the 90th and 95th percentiles as alternative measures of inequality is motivated by recent studies documenting that income inequality dynamics are strictly related to top end dynamics \parencite{10.1093/qje/qjx043}.} and alternative samples (excluding the covid-19 period and omitting countries for which the financial shock series is particularly volatile) are summarized in the robustness checks section of Appendix (\ref{app_robustness}).

 \subsection{Segments of the income distribution}\label{interquintile_results}
 
\begin{figure}[h!]
\centering
\begin{minipage}{0.32\textwidth}
  \centering
  \includegraphics[width=\linewidth]{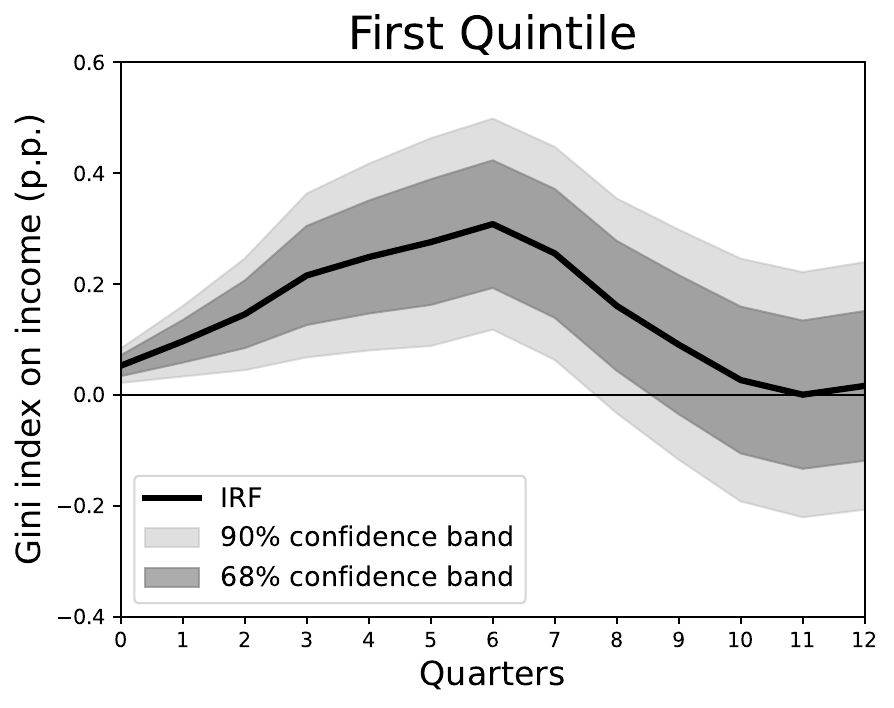}
\end{minipage}\hfill
\begin{minipage}{0.32\textwidth}
  \centering
  \includegraphics[width=\linewidth]{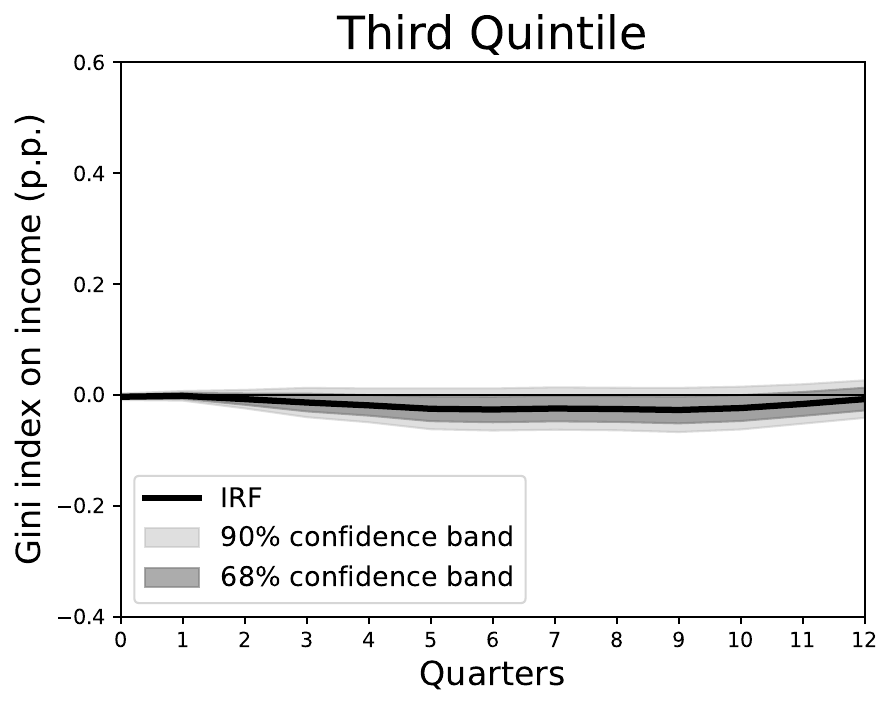}
\end{minipage}\hfill
\begin{minipage}{0.32\textwidth}
  \centering
  \includegraphics[width=\linewidth]{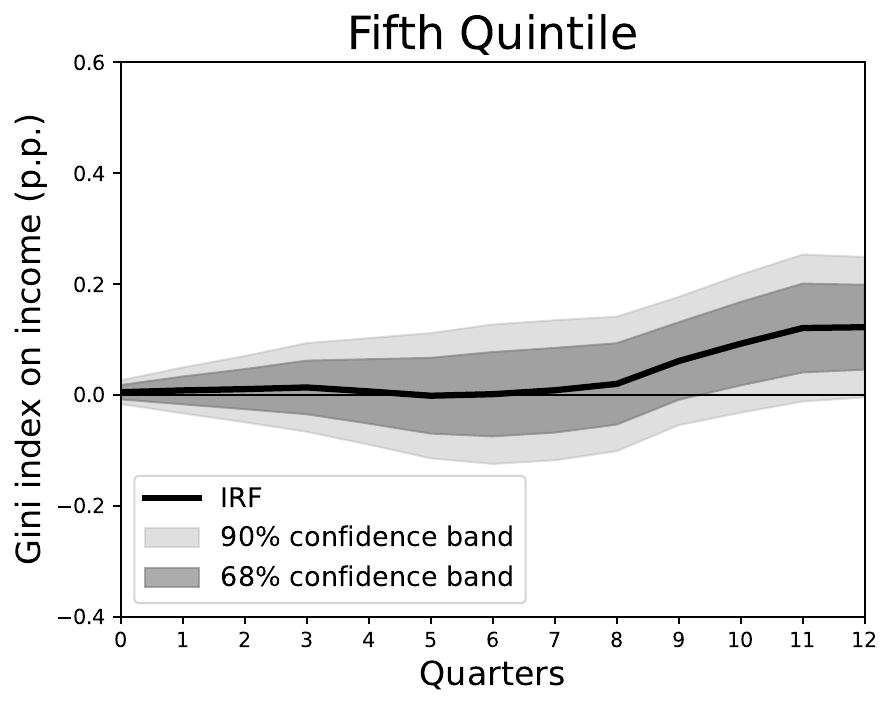}
\end{minipage}
\caption{Impulse responses to financial shocks: interquintile inequality}
 \vspace{0.15cm}
\footnotesize
The figure displays the impulse responses of interquintile inequality for each income quintile following a one-standard-deviation financial shock. Solid lines represent point estimates; shaded areas denote 90\% and 68\% confidence bands computed from panel HAC standard errors.
    \label{fig:income_segments_irf}
\end{figure}

Do different segments of the income distribution respond differently to a financial shock? Figure \ref{fig:income_segments_irf} presents the impulse responses of inequality across specific segments of the distribution.\footnote{For space considerations, we report only the first, third, and fifth quintiles. The responses of the second and fourth quintiles closely resemble those of the third.} Specifically, we consider interquintile Gini indices.\footnote{To gain insight into both the variability and the average effects, we also examined the responses of average income by quintile. For visibility purposes, the results are available upon request.} This approach enables us to identify where in the income distribution the effects are most pronounced. 

Following a positive financial shock, we observe heterogeneous effects across the income distribution. Interquintile Gini indices more significantly in the first and fifth quintiles. In the bottom quintile (first), the response is characterized by a temporary increase in inequality. In contrast, the top quintile (fifth) experience an increase in the Gini index at longer horizons. Overall, the tails of the income distribution are more affected by financial shocks than middle-income counterparts.\\
This result aligns with previous research, which, although it did not directly address financial shocks, has shown that the tails of the income distribution are more sensitive to macroeconomic shocks than middle-income groups. For instance, \textcite{amberg2022five} find that monetary policy shocks generate a U-shaped pattern in the total income response, meaning they have significantly larger effects on incomes at the tails of the distribution compared to the middle-income groups.\\
Concerning the mechanisms behind these distributional patterns, the response at the top of the income distribution is likely due to differences in financial asset holdings and to the role of the skill composition in shaping labor market outcomes. In contrast, the dynamics at the bottom of the distribution might reflect the specific composition of individuals in this segment. In particular, we are looking at very low income levels, characterized by low elasticity to financial shocks. However, some of these individuals may still be indirectly affected by such shocks. For instance, part of their income may stem from precarious or marginal jobs loosely tied to financial or real estate market activity, which tend to be sensitive to financial conditions. In such cases, even in the absence of job losses, a compression of already low incomes may occur. Clearly, individuals within the first quintile are heterogeneous, and some may still benefit—albeit marginally—from the shock, for example, through intergenerational transfers or financial income held by other household members. These internal differences can lead to a widening of income dispersion within the quintile. We directly explore the empirical relevance of these mechanisms in the next section. 

\subsection{The role of financial and labor income components}\label{res_fin_lab}
 
The results from the previous section show that positive financial shocks increase income inequality, and the distributional effects are more pronounced at the tails of the income distribution.\\
To understand the transmission mechanics behind such a heterogeneous response, it is necessary to disentangle the relative contributions of different income components. Our data enable us to decompose individual income into financial and labor income components, allowing us to quantify the responses of financial and labor income inequality to the financial shock. Notably, while the distinction between financial and labor income channels has gained increasing attention in the study of macroeconomic shocks—particularly in the monetary policy literature (see, e.g., \textcite{amberg2022five, andersen2023monetary})—it remains largely unexplored in the context of financial shocks.

\begin{figure}[htbp]
    \centering
    \begin{minipage}{0.40\textwidth}
        \centering
        \includegraphics[width=\linewidth]{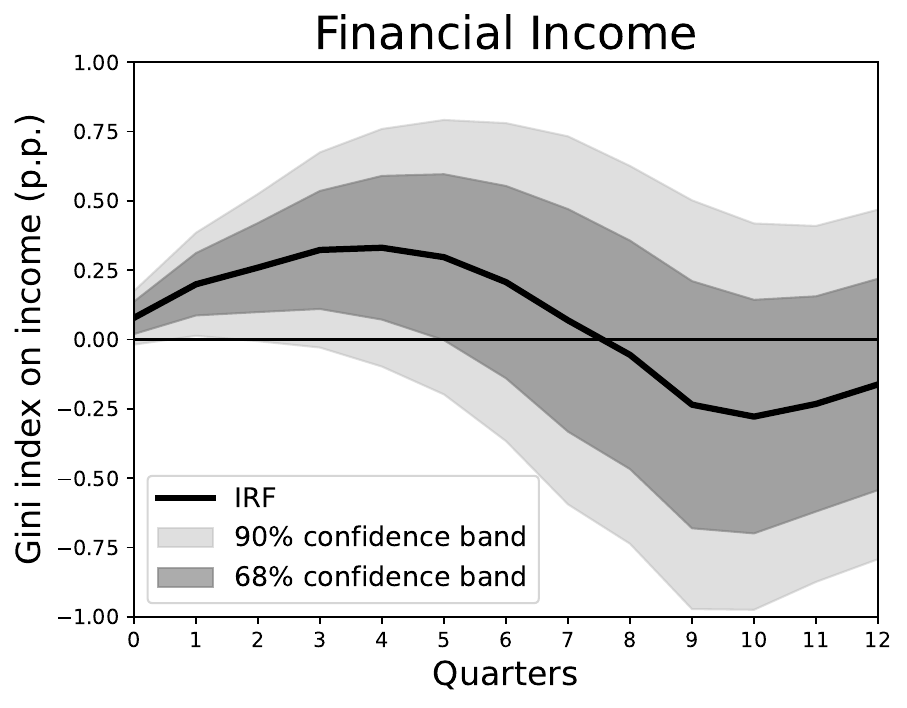}
    \end{minipage}
    \hspace{0.1cm}
    \begin{minipage}{0.40\textwidth}
        \centering
        \includegraphics[width=\linewidth]{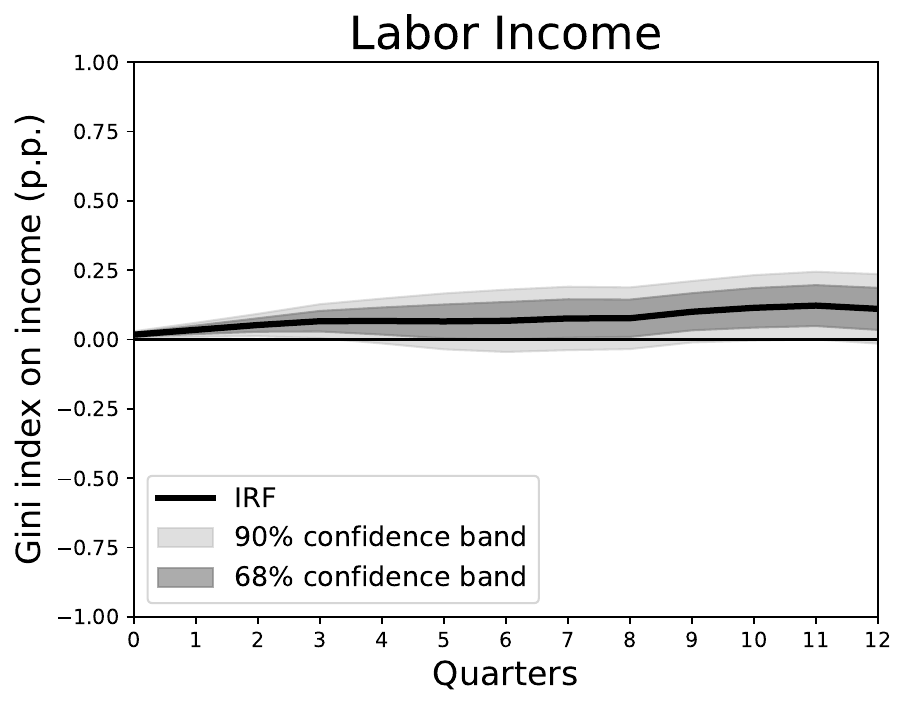}
    \end{minipage}
    \caption{Financial (left) and labor (right) income inequality responses}
    \vspace{0.15cm}
\footnotesize
The figure illustrates the response of inequality for financial and labor incomes following a one-standard-deviation positive financial shock. Full lines are point estimates; shaded areas indicate 90\% and 68\% confidence bands computed from panel HAC standard errors.
    \label{fig:fin_and_lab_income_inequality}
\end{figure}

\begin{figure}
    \centering
    \includegraphics[width=0.65\linewidth]{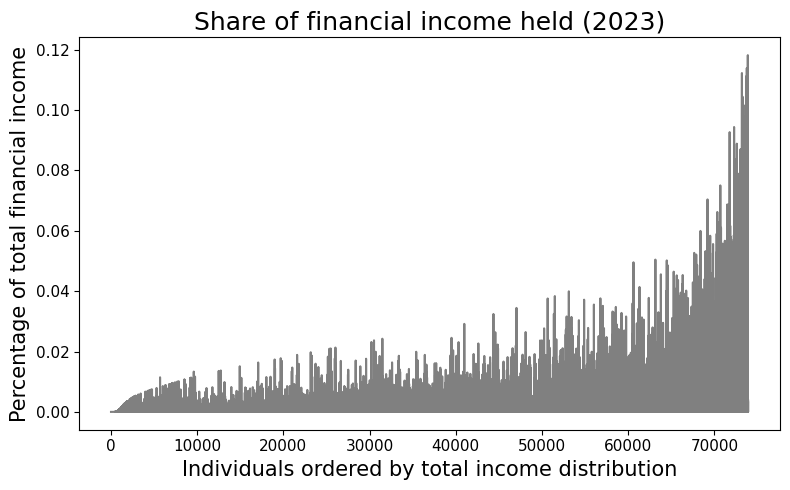}
    \caption{Financial income distribution in the Euro Area (2023)}
    \vspace{0.15cm}
\footnotesize
The figure displays the share of total financial income held by each individual, with individuals ranked according to their position in the overall income distribution. Top earners (in the 99th percentile) are excluded from the graphical representation for clarity and visibility purposes.
    \label{fig:financial_income_holdings}
\end{figure}

Figure \ref{fig:fin_and_lab_income_inequality} provides evidence on the relevance of financial shocks for both income components. The left panel displays the impulse response of financial income inequality, while the right panel shows the response of labor income inequality.\\
Following a positive financial shock, financial income inequality rises, peaking at approximately 0.32 percentage points after one year, before gradually declining at longer horizons. By contrast, the response of labor income is less immediate but equally significant: labor income inequality persists over time, stabilizing at around 0.12 percentage points after two years.\\
The first result, concerning financial income, appears straightforward and intuitive. It might reflect the distribution of financial asset holdings: when a positive financial shock occurs, share prices rise, leading to higher returns for individuals with financial income—returns that are larger for those with greater financial holdings. In this sense, the financial shock acts as an amplifier of the pre-existing uneven distribution of financial assets along the income distribution. To provide supporting evidence for this mechanism, Figure \ref{fig:financial_income_holdings} displays the share of financial income held across income distribution in our sample.\footnote{Using a consistent sample, we compare the distribution of financial income in the Euro Area for a specific year based on EU-SILC and ECB’s Household Finance and Consumption Survey (HFCS) data, in order to assess potential underreporting of financial income in our sample. This concern has been raised in the context of United States survey data, where studies \parencite{brady2021comparing} have documented that financial income is underestimated. The relevant figures are in Appendix \ref{comparison}. The two distributions have a similar shape: Both are highly right-skewed.}

Since financial income captures only individuals who actively participate in financial markets, this result complements the broader literature on financial development and income inequality (see, e.g., \cite{de2017finance}). Specifically, it aligns with the notion of the intensive margin emphasized in previous studies (\cite{greenwood1990financial}). While much of the existing literature focuses on financial development — understood as the expansion of financial activity — we interpret our findings as consistent with the view that positive financial shocks benefit those already integrated into financial markets.\\
In this context, the results in \textcite{gocmen2025private} are not only complementary to our findings but also an integral part of the broader narrative. The complementarity element stands out in emphasizing how portfolio allocation decisions (i.e., financial income allocation) shape individual income trajectories and, consequently, influence distributional outcomes. Furthermore, the results in \textcite{gocmen2025private} extend the comprehension of the mechanism we document: while we show that a positive financial shock raises share prices and amplifies financial income inequality, \textcite{gocmen2025private} find, using United States data, that differences in portfolio composition are crucial. Specifically, they find that wealthier individuals have access to more profitable investment opportunities, thereby reinforcing inequality through returns heterogeneity.\\
A less immediate result concerns labor income inequality. First, labor income responds to financial shocks; second, its response appears to be persistent over time. We interpret this as related to the type of occupations held by workers. The suggested underlying mechanism is that a positive financial shock, by boosting share prices, disproportionately benefits the compensation of individuals in top occupations or those who hold equity stakes in their companies. To provide some rough evidence for this channel, we rely on the ISCO-08 occupational classification developed by the ILO, which categorizes workers based on their job tasks, thereby allowing us to distinguish between high-skilled and low-skilled workers. The fundamental conjecture here is that companies' equity stakes are held mainly by high-skilled workers. This mechanism would be coherent with \textcite{adam2016firms}’s idea that the manner in which firms reward individuals for their labor contributes significantly to observed levels of income inequality.

\begin{figure}
    \centering
    \includegraphics[width=0.5\linewidth]{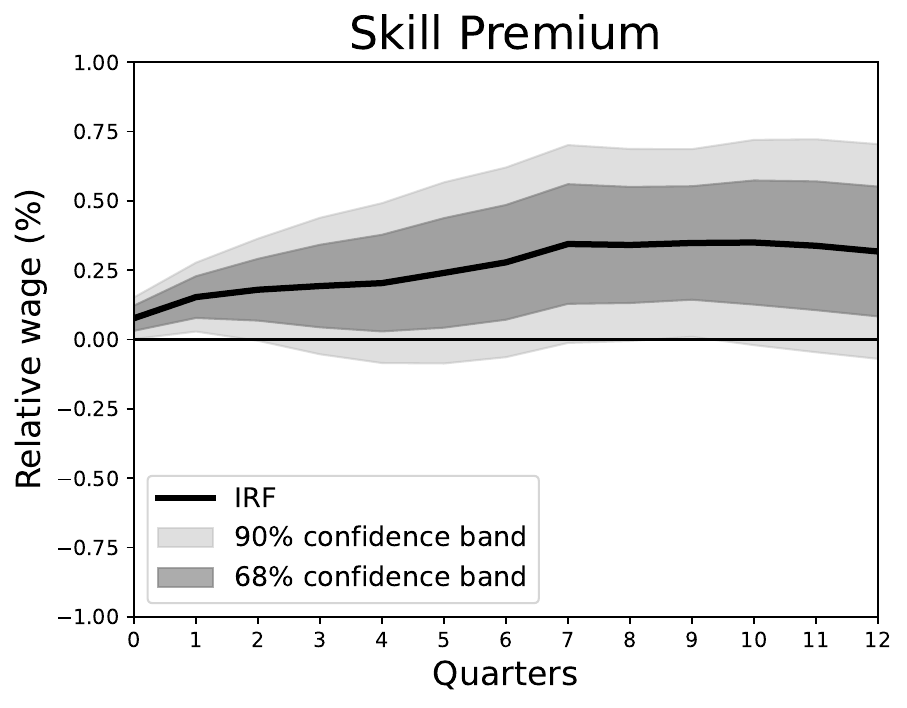}
    \caption{The effects of financial shocks on the skill premium}
    \vspace{0.15cm}
\footnotesize
The figure shows the response of the skill premium to a one-standard-deviation positive financial shock. The solid line represents point estimates, while the shaded areas denote 90\% and 68\% confidence bands computed from panel HAC standard errors.
    \label{fig:skill_premium}
\end{figure}

Figure \ref{fig:skill_premium} presents the estimated effects of a financial shock on the skill premium, defined as the log ratio between high-skilled and low-skilled workers' wages. The results provide supporting evidence for the proposed mechanism.\footnote{Results for high-skilled and low-skilled workers are presented in Figure \ref{fig:skill_vs_low_skill_spec} (Appendix \ref{additional_results}). Following a positive financial shock, wages increase for high-skilled workers, but not for low-skilled workers. Additional evidence on hours worked reveals unclear variations, suggesting that the observed effects are attributable to changes in compensation rather than to alternative explanations, such as technological displacement.} We interpret this result as complementary to recent macroeconomic evidence. While the mechanism is novel in the context of financial shocks, the notion that aggregate shocks can have unequal effects across skill groups is consistent with prior research. For example, \textcite{dolado2021monetary} show that unexpected monetary policy easing widens inequality between high- and low-skilled workers.

\section{The asymmetric response of inequality to financial shocks}\label{sign_effect}

In this section, we examine the potential asymmetric effects of positive and negative financial shocks on income inequality. This focus is motivated by recent macroeconomic literature highlighting the importance of such nonlinearities with respect to financial shocks (e.g. \cite{brunnermeier2014macroeconomic, barnichon2022effects}) and by the policy relevance of understanding the distinct impact of negative shocks.\\
There are at least two reasons—although not necessarily exhaustive—why adverse financial shocks are of particular concern for policymakers. First, they pose a serious threat to the financial stability. A negative financial shock typically depresses asset prices, increases banking sector stress, and can trigger a credit crunch, thereby evolving into a systemic risk for financial stability. Second, financial market fluctuations have substantial implications for the real economy. A large body of literature has documented the macroeconomic effects of financial shocks (i.a. \textcite{gilchrist2009credit, peersman2011bank, peersman2014shocks, caldara2016macroeconomic, furlanetto2019identification}). In both cases—threats to financial stability and to the real economy—there might be arguments for policy intervention, whether by financial stability authorities or by national governments through macroprudential policies or fiscal stimulus plans. Moreover, in contrast to positive financial shocks—where policymakers might face a trade-off between stimulating investment and containing inequality—negative shocks naturally prompt intervention to stabilize the economy, thereby offering an opportunity to design policies that simultaneously address both macroeconomic stabilization and distributional concerns.
 
\subsection{Separating positive and negative financial shocks}

To distinguish between positive and negative financial shocks, we draw on the idea that both have similar implications for second moments. This concept connects to the financial literature on sign asymmetries. For instance, \textcite{ferreira2007asymmetric} build on the fact that negative shocks to stock prices tend to generate larger increases in volatility than positive shocks of similar magnitude. A related notion is adopted by \textcite{forni2017news}, who, despite focusing on the relationship between news and uncertainty, highlight the existence of asymmetric effects on uncertainty—conceptually close to the notion of volatility.\\
Building on these notions, we extend the baseline identification scheme by introducing an additional variable to explicitly capture financial market volatility, following an approach similar to that adopted by \textcite{lucidi2024effects}, who separate negative and positive temperature shocks trough the variance of the vapor pressure. We approximate the financial market volatility with the ECB's CLIFS. We impose the condition that both positive and negative financial shocks affect financial market volatility positively on impact, thereby distinguishing between the two types of shocks. While we remain agnostic about the relative magnitude of their effects on the variance, the impulse responses show that the impact response of financial volatility is approximately twice as large following a negative financial shock, consistent with the findings in the financial literature.\footnote{To save space and for clarity of exposition, the results from the first step estimates are available upon request.} The updated identification scheme is summarized in Table \ref{tab:restrictions_id2} of the Appendix (\ref{app_nonlin}). The reason why we do not consider a more standard approach, tackling this nonlinearity directly in the LP,\footnote{The literature has increasingly focused on how to best model various forms of nonlinearities, with particular attention to sign asymmetries. Other notable contributions in this area include \textcite{ben2019identification} and \textcite{caravello2024disentangling}.} is that by introducing the sign nonlinearity in the P-LP only, we would have identified the shock in a linear P-SVAR framework and then estimated its effects nonlinearly, which may not be methodologically sound, as pointed-out in \textcite{barnichon2022effects}.\footnote{Since we identify the financial shock without considering the outcome variable in the P-SVAR (and hence no linear or nonlinear relationship is assumed between the two), we still use this alternative nonlinear specification as a robustness check. The specification in this case is (\textcite{tenreyro}): 
\begin{equation}
y_{i,t+h} = \beta^{+}_h\max[0, \varepsilon_{i,t}] + \beta^{-}_h\min[0, \varepsilon_{i,t}] + \sum_{l=1}^{p}{\gamma_{h,l} X_{i, t-l}} + \sum_{l=0}^{p}{\psi_{h,l} U_{i, t-l}} + \vartheta \bar{x}_i + \rho \bar{x}_t + v_{i, t+h},
\end{equation}
where \(\beta^{+}_h\) and \(\beta^{-}_h\) are the coefficients of interest for the positive and negative shocks, respectively.\\The asymmetry is robust to the alternative specification. Regarding the separate responses, the positive shock effect is confirmed under this specification, while the response to a negative shock remains insignificant.}

\subsection{Results}

\begin{figure}[htbp]
    \centering
    \begin{subfigure}[t]{0.35\textwidth}
        \centering
        \includegraphics[width=\linewidth]{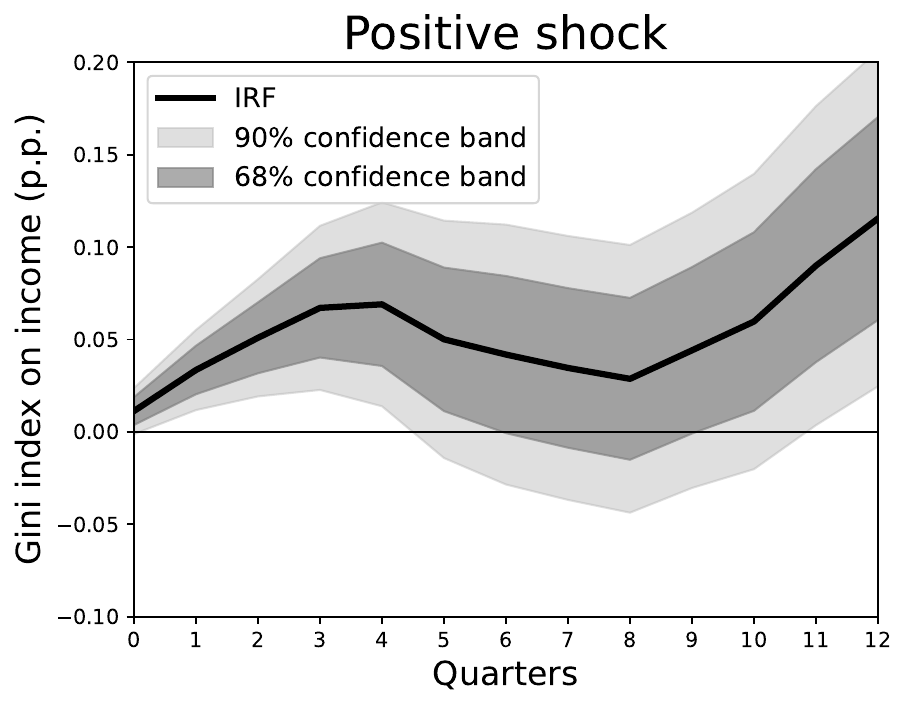}
        \label{fig:first_image}
    \end{subfigure}
    \hspace{0.1cm}
    \begin{subfigure}[t]{0.35\textwidth}
        \centering
        \includegraphics[width=\linewidth]{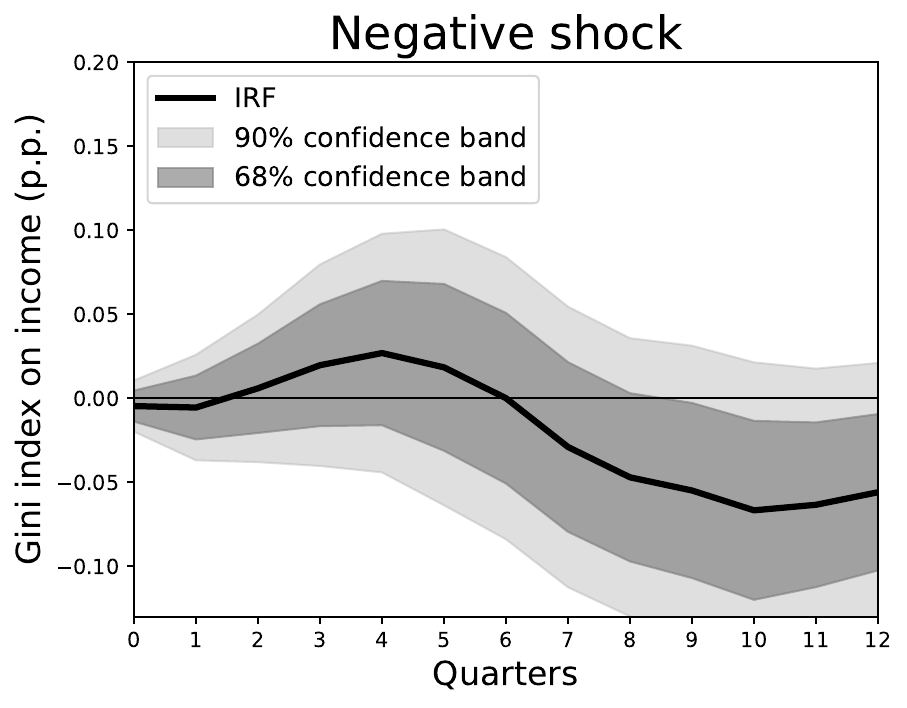}
        \label{fig:second_image}
    \end{subfigure}
    \caption{The asymmetric distributional effects of financial shocks}
    \vspace{0.15cm}
\footnotesize 
The figure shows the response of income inequality to a one-standard-deviation positive (left) and negative (right) financial shock. The solid line represents point estimates, while the shaded areas denote 90\% and 68\% confidence bands computed from panel HAC standard errors.
    \label{fig:size_effect}
\end{figure}

Figure \ref{fig:size_effect} presents the estimated effects of positive (left panel) and negative (right panel) financial shocks on income inequality in the Euro Area. Following a positive financial shock, inequality rises, suggesting a permanent increase in inequality, consistent with the undistinguished positive shock dynamics displayed in Figure \ref{fig:gini_euro_aggregate}. In contrast, negative financial shocks generate an initially unclear response that turns negative over longer horizons. This pattern underscores the presence of pronounced asymmetries in the distributional effects of financial shocks.\\
Such a "sign effect" has been emphasized theoretically by \textcite{brunnermeier2014macroeconomic} and documented empirically by \textcite{barnichon2022effects}. Our findings confirm this asymmetry, albeit in the context of income inequality. The main differences between positive and negative financial shocks lie in both the sign and the timing of their effects on inequality. Regarding the sign, this asymmetry is consistent with the operation of the previously discussed “inequality amplifier” mechanism. A positive financial shock tends to magnify existing disparities over time by reinforcing cumulative gains among high-income and high-wealth individuals. In contrast, a negative financial shock leads to a reduction in inequality, as wealthier households typically experience proportionally larger losses.\\
In terms of dynamics, the persistent decline in inequality observed at longer horizons likely reflects the slow adjustment process that follows financial downturns. After a market collapse, the most exposed agents suffer losses that are not easily reversible, requiring a longer period to rebuild their portfolios. This persistence may also be associated with deleveraging phases and heightened risk aversion that typically follow adverse financial episodes, both contributing to a protracted redistribution of wealth and income. Finally, the lagged adjustment suggests that the redistributive effects of financial downturns may also reflect the sluggish response of labor market conditions.

\begin{figure}[htbp]
    \centering
    \begin{subfigure}[t]{0.33\textwidth}
        \includegraphics[width=\linewidth]{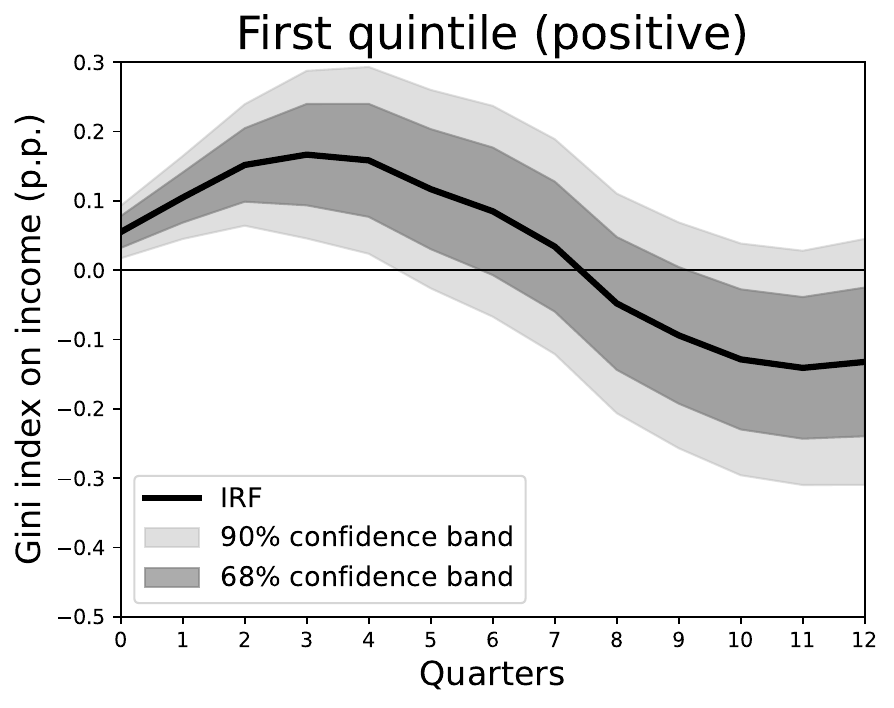}
    \end{subfigure}
    \hspace{0.1cm}
    \begin{subfigure}[t]{0.33\textwidth}
        \includegraphics[width=\linewidth]{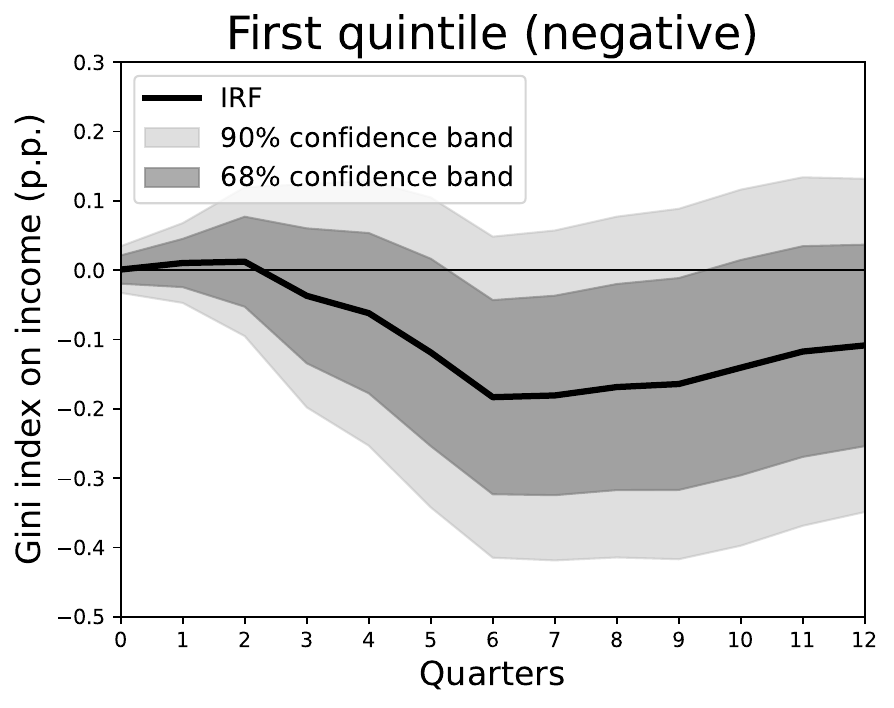}
    \end{subfigure}

    \vspace{0.2cm}

    \begin{subfigure}[t]{0.33\textwidth}
        \includegraphics[width=\linewidth]{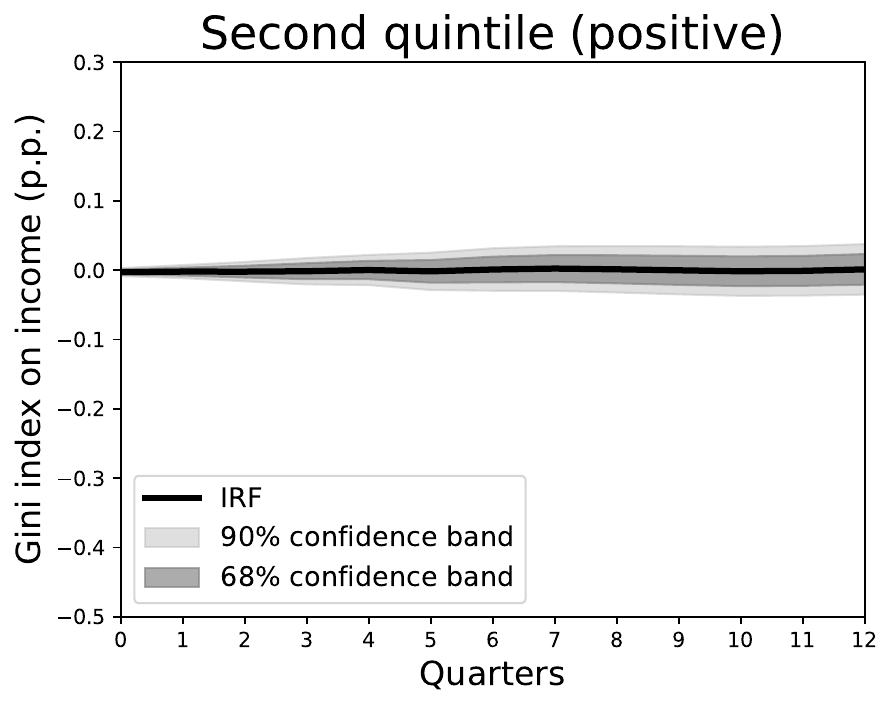}
    \end{subfigure}
    \hspace{0.1cm}
    \begin{subfigure}[t]{0.33\textwidth}
        \includegraphics[width=\linewidth]{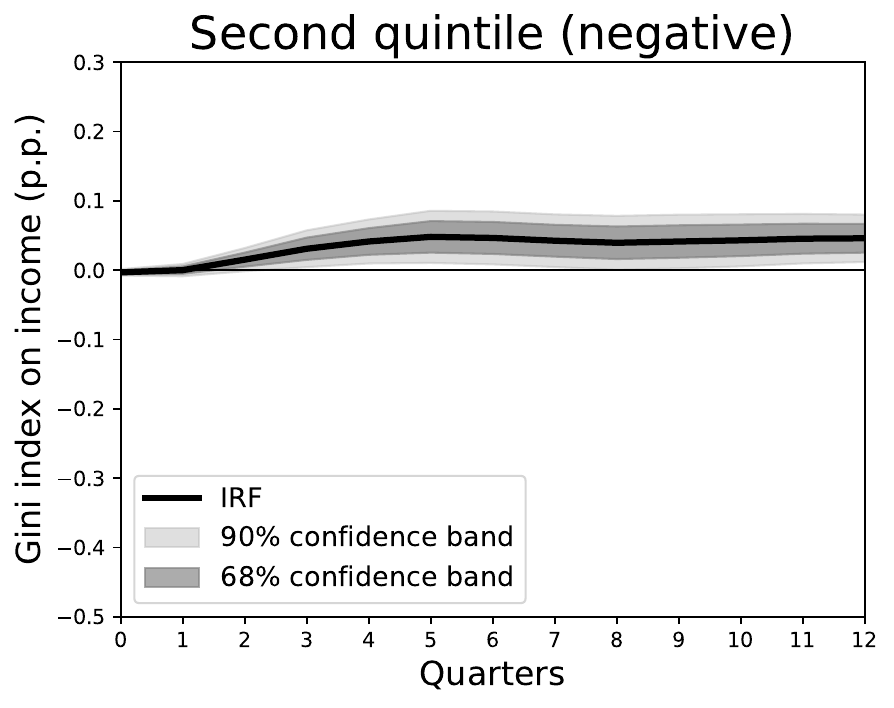}
    \end{subfigure}

    \vspace{0.2cm}

    \begin{subfigure}[t]{0.33\textwidth}
        \includegraphics[width=\linewidth]{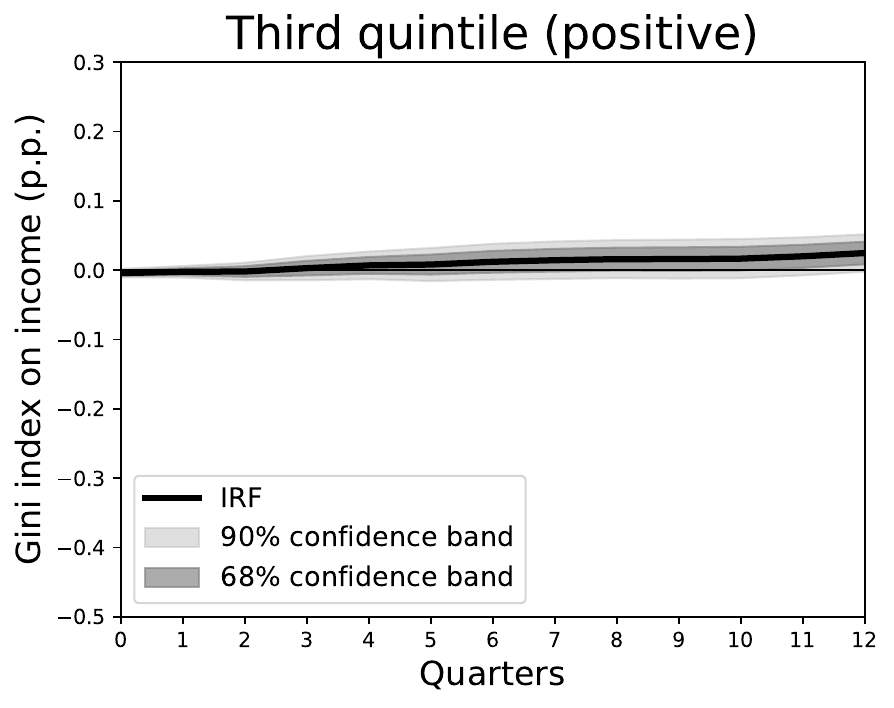}
    \end{subfigure}
    \hspace{0.1cm}
    \begin{subfigure}[t]{0.33\textwidth}
        \includegraphics[width=\linewidth]{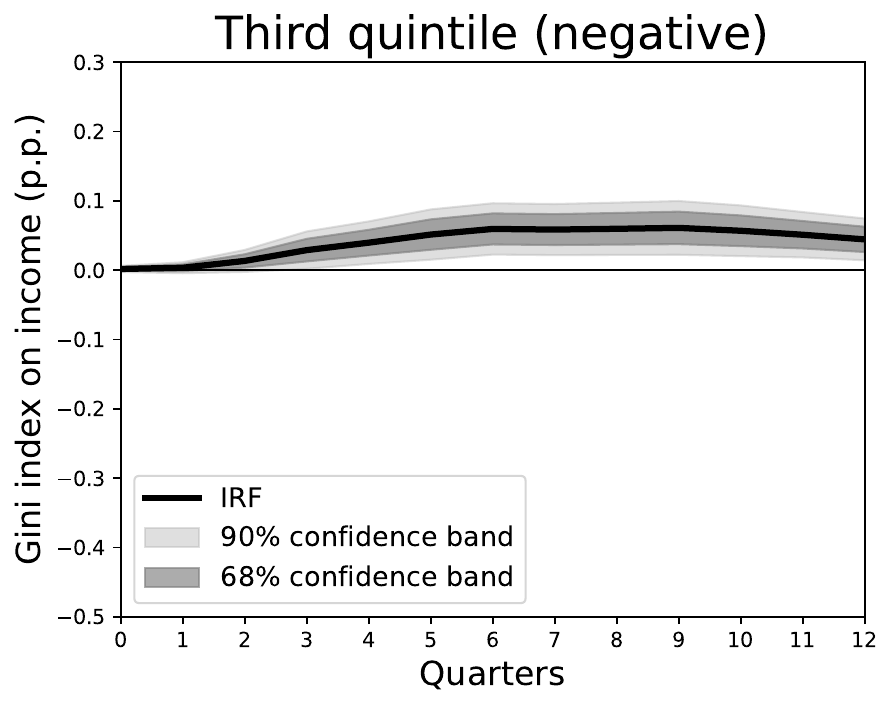}
    \end{subfigure}

    \vspace{0.2cm}

    \begin{subfigure}[t]{0.33\textwidth}
        \includegraphics[width=\linewidth]{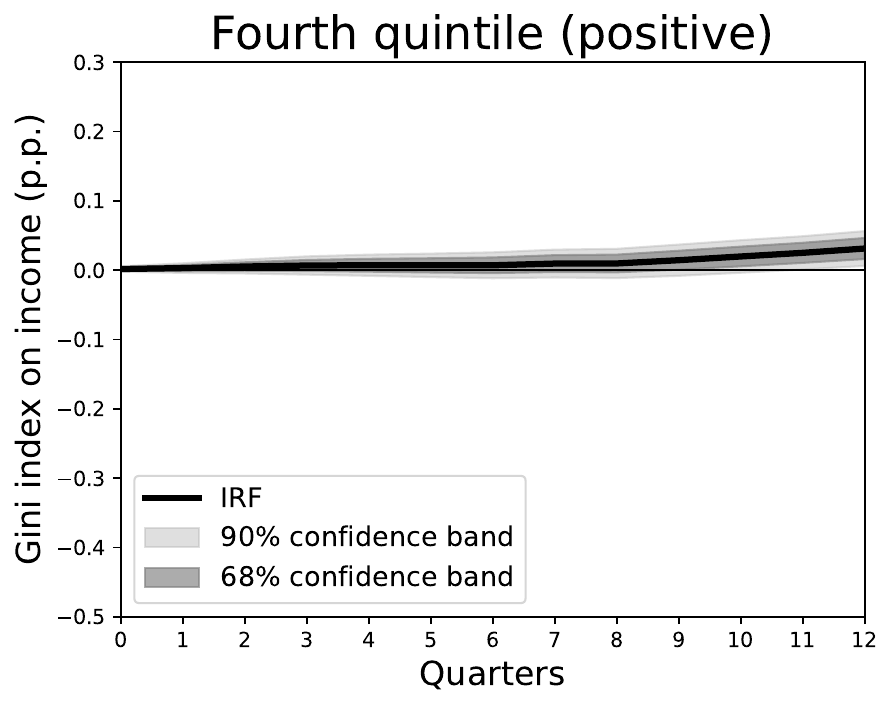}
    \end{subfigure}
    \hspace{0.1cm}
    \begin{subfigure}[t]{0.33\textwidth}
        \includegraphics[width=\linewidth]{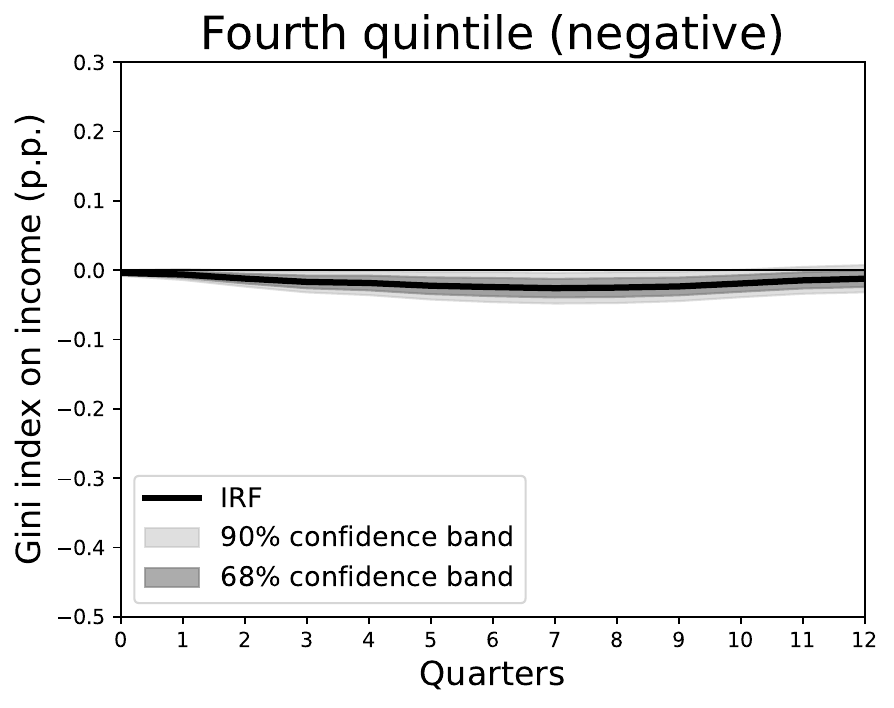}
    \end{subfigure}

    \vspace{0.2cm}

    \begin{subfigure}[t]{0.33\textwidth}
        \includegraphics[width=\linewidth]{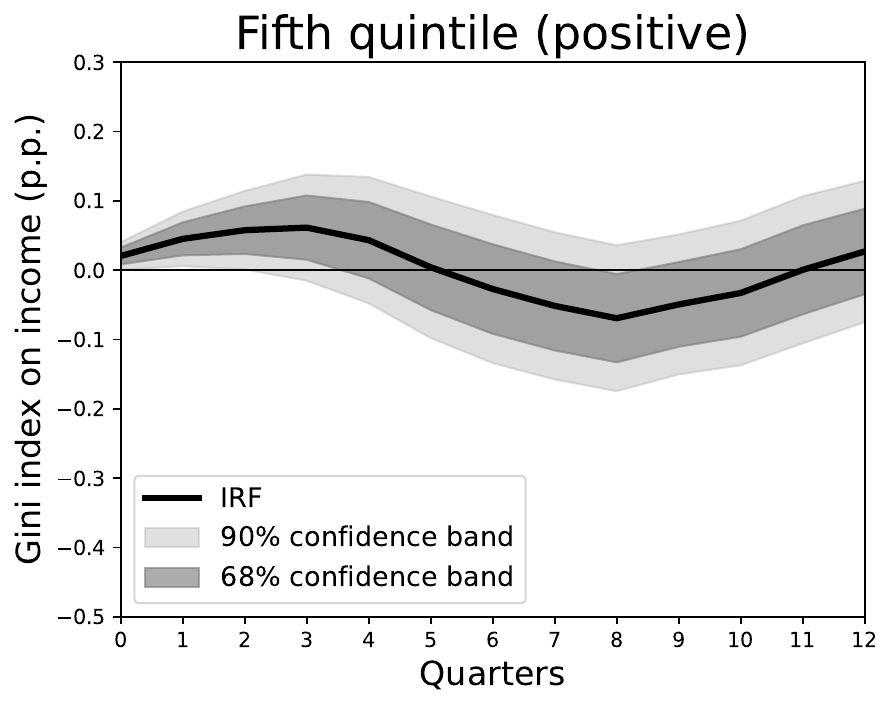}
    \end{subfigure}
    \hspace{0.1cm}
    \begin{subfigure}[t]{0.33\textwidth}
        \includegraphics[width=\linewidth]{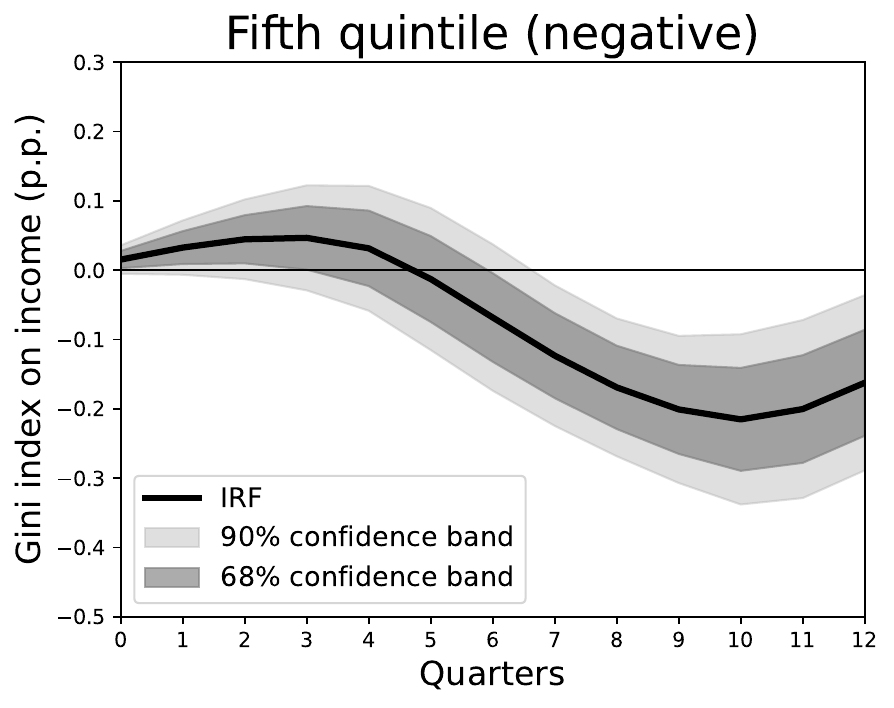}
    \end{subfigure}

    \caption{IRFs across quintiles: positive (left) vs. negative Shocks (right)}
    \vspace{0.15cm}
\footnotesize 
The figure shows the response of income inequality in different quintiles to a one-standard-deviation positive (left column) and negative (right column) financial shock. The solid line represents point estimates, while the shaded areas denote 90\% and 68\% confidence bands computed from panel HAC standard errors.
    \label{fig:quintile_shocks_sign}
\end{figure}

Turning to the heterogeneity across income segments (Figure \ref{fig:quintile_shocks_sign}) we find that after positive financial shocks, the tails of the income distribution respond more strongly, consistent with our earlier evidence. In particular, the dynamics for the top quintile resemble the aggregate pattern shown in Figure \ref{fig:size_effect}, though with a less pronounced amplitude — inequality initially rises, then declines before increasing again. Adverse financial shocks reveal new and asymmetric patterns. First, income inequality within the top quintile decreases significantly, consistent with the idea that top-income individuals are more exposed to financial market losses. Here, the inequality amplification mechanism previously documented works in reverse: a negative shock compresses relative distances within the quintile. Second, we find that inequality increases within middle-income groups (second and third quintiles). Credit market frictions likely drive this result: when financial conditions tighten, households that rely more heavily on credit see their disposable income squeezed, while others remain less affected. Sectoral differences may also play a role, as some jobs are more sensitive to financial shocks than others. In addition, the structure of household income, such as dual earners or access to financial buffers, can further amplify intra-group dispersion. Lastly, the absence of a positive and significant response in the bottom quintile may reflect both the limited exposure to financial markets and the presence of collective bargaining or minimum income schemes, which help stabilize low-end earnings. This latter result aligns with the insights of \textcite{heathcote2010unequal}, who emphasize the role of government redistribution in protecting the lower end of the distribution. We interpret the result for the first quintile as complementary to this vision. Moreover, our findings are broadly consistent with the evidence provided by \textcite{gornemann2021doves}, who show that income composition is a key predictor of which households benefit or lose following aggregate shocks. This finding is particularly relevant to our results, as it highlights how differences in household balance sheets contribute to explaining why middle-income groups are responsive to such shocks.

 \newpage
 \section{Conclusions}\label{conclusions}

 This paper has examined the effects of financial shocks on income inequality in the Euro Area, contributing to a growing literature at the intersection of macroeconomics, finance, and inequality. Using a two-step empirical approach combining structural panel vector autoregressions and panel local projections, we document that financial shocks increase income inequality, with substantial heterogeneity across income quintiles. Much of the increase in inequality is explained by the responsiveness observed at the tails of the distribution. By decomposing income into labor and financial components, we uncover distinct distributional channels: financial shocks amplify inequality through the intensive margin of financial income, while also increasing labor income inequality via higher compensation at the top of the income distribution. These findings highlight the varied transmission mechanisms linking financial shocks to inequality, underscoring the importance of considering both income composition and labor heterogeneity.\\
We find that adverse financial shocks have a larger impact on income inequality than positive shocks, even if the persistence profile of the effects indicates that positive shocks trigger longer-lasting inequality effects. Notably, adverse financial shocks significantly affect the distributional pattern of middle-income groups, highlighting differentiated vulnerabilities across the income distribution. From a policy perspective, our results suggest that financial shocks exacerbate existing inequalities, raising important considerations for the design of financial stability, redistribution, and social protection policies.\\
While our analysis focuses on intensive margins, an important direction for future research is to investigate the role of extensive margins in shaping the inequality effects of financial shocks—both through access to financial income and through labor market participation, particularly in relation to unemployment dynamics. Similarly, investigating how differences in portfolio allocation and wealth inequality mediate the transmission of financial shocks could provide further insights into the distributional consequences of financial instability. 

\newpage

\printbibliography
\clearpage
\appendix
\section*{Appendix}
\addcontentsline{toc}{section}{Appendix}
%\pretocmd{\figure}{\begingroup\setstretch{1.0}}{}{}
%\apptocmd{\endfigure}{\endgroup}{}{}

\section{Data} \label{data_app}

\textbf{GDP} is log of real GDP obtained from the European Union Statistical Office (Eurostat) Database; \textbf{Prices} is log GDP deflator from Eurostat; \textbf{Investment/output} is log of investments (Eurostat) over GDP;  \textbf{Stock prices} is log of share prices index from the Organisation for Economic Co-Operation and Development (OECD) Database; \textbf{Interest rate} is the short-term interest rate from OECD; \textbf{Gini index} is Gini coefficient of gross market income (percentage) computed from European Union Statistics on Income and Living Conditions (EU-SILC) survey; \textbf{P90} is the 90th income percentile computed from EU-SILC; \textbf{P95} is the 95th income percentile computed from EU-SILC; \textbf{Financial deepening} is the ratio between financial assets and GDP expressed in percentage points (Eurostat); \textbf{CLIFS} is the Country-Level Index of Financial Stress from the European Central Bank (ECB); \textbf{Skill premium} is defined as the wage ratio between high-skilled and low-skilled workers in working age, and is constructed using data from EU-SILC; \textbf{Uncertainity index} is the smoothed version of the World Uncertainity Index by \cite{ahir2022world}.\footnote{The Uncertainty Index was not available for Estonia and Luxembourg. For Estonia, we imputed the value using the average of the available data for the other Baltic countries (Latvia and Lithuania); for Luxembourg, we used the average of Belgium and the Netherlands. The same approach was applied to impute the missing CLIFS data for Estonia.}\\
Data are collected on a quarterly basis. For the outcome variables (Gini index, P90, and P95), which are only available at annual frequency, we apply a standard linear interpolation to get the quarterly figures. This imputation method assumes a uniform quarterly rate of change between annual data points, which aligns with the slow-moving nature of the variables, where intra-annual fluctuations are less common.\footnote{Figure \ref{fig:alt_trim_gini} in the Appendix shows that the increase in income inequality following a financial shock is robust to the method used for converting data to quarterly frequency. As a robustness check, we employ a flat interpolation method, assigning the annual value uniformly across all quarters.} The CLIFS variable is available at the monthly frequency; to align it with the quarterly frequency of the other series, we compute the quarterly average. GDP, Prices, Investment/output, and Stock prices are multiplied by 100.

\section{PVAR Specification} \label{bayesian_spec}

Our panel VAR specification follows a pooled panel approach, meaning that the only panel feature is that the dataset comprises observations coming from different units (countries). For clarity and consistency with standard practice, we present the VAR directly in its vectorized form: 

\begin{equation}
    y = \bar{x}\beta+\epsilon
\end{equation}
The parameters of interest are $\beta$ and $\Sigma_c$ (the unit residual covariance matrix), the former including section-specific constants. Given that the structure is essentially that of a VAR, standard Bayesian estimation techniques for the posterior distribution can be applied. We adopt a traditional Normal–Wishart prior setup, in which $\beta$ is assumed to follow a multivariate normal distribution and $\Sigma_c$ an inverse Wishart distribution, resulting in corresponding posterior distributions. Parameter values follow the conventional choices documented in \textcite{dieppe2018bayesian}.\\
The implementation of sign restrictions follows the methodology developed by \textcite{arias2014inference}, and draws are generated through Gibbs sampling. We use a total of 2,000 iterations, with a burn-in of 1,000.\\
Under these settings, the estimation of the baseline model (Table \ref{tab:restrictions_id1}) completes in approximately one hour, while the nonlinear extension (Table \ref{tab:restrictions_id2}) requires around five days.\footnote{Computations are performed on a machine equipped with an Intel(R) Xeon(R) W5-3435X 3.10 GHz processor and 64 GB of RAM.}

\newpage

\section{Additional results} \label{additional_results}

\begin{figure}[H]
    \centering
    \includegraphics[width=0.45\linewidth]{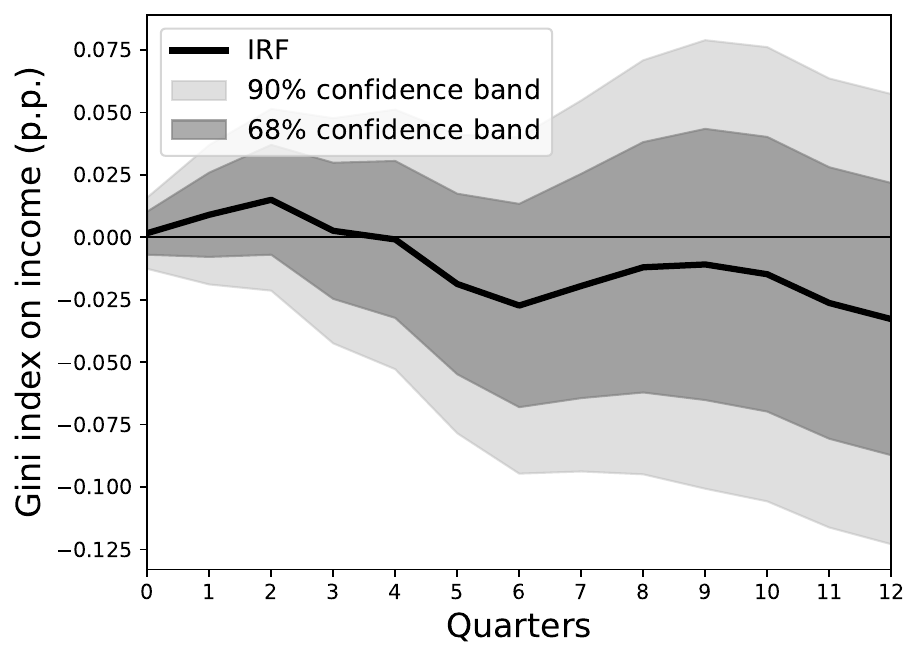}
    \caption{The effects of demand shocks on income inequality}
    \vspace{0.15cm}
\footnotesize 
The figure shows the response of income inequality after a one-standard-deviation positive generic demand shock occurs. Full lines are point estimates; shaded areas indicate 90\% and 68\% confidence bands computed from panel HAC standard errors.
\label{fig:inequ_dem_shock}
\end{figure}

\begin{figure}[H]
    \centering
    \begin{subfigure}[t]{0.33\textwidth}
        \includegraphics[width=\linewidth]{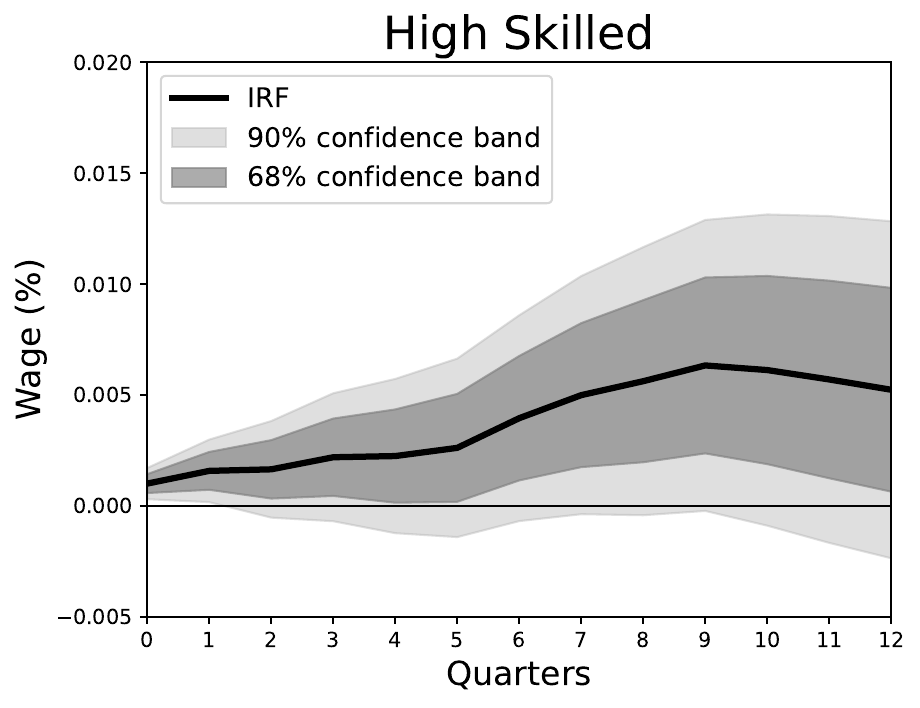}
    \end{subfigure}
    \hspace{0.1cm}
    \begin{subfigure}[t]{0.33\textwidth}
        \includegraphics[width=\linewidth]{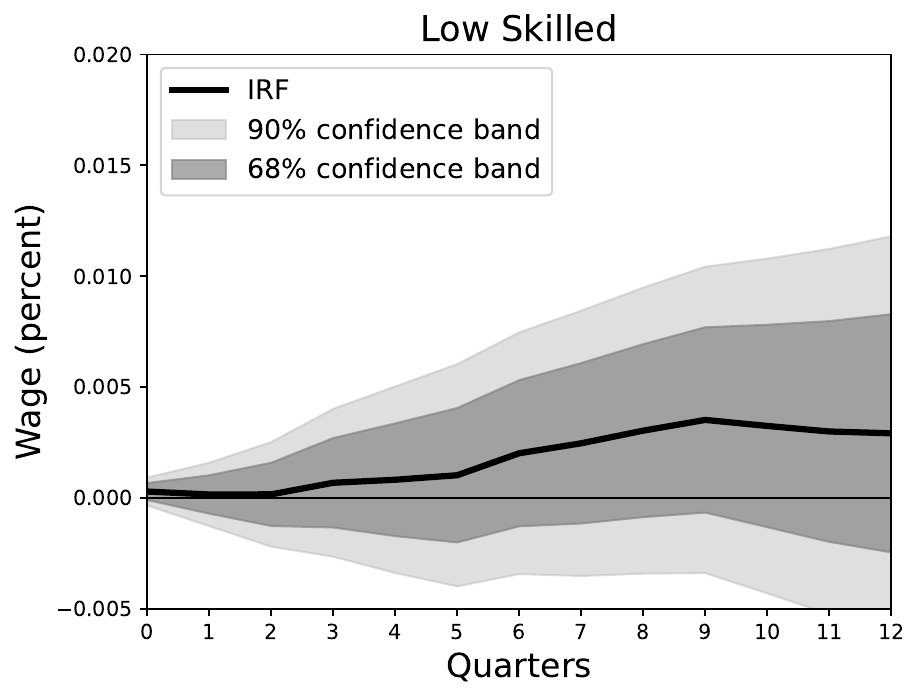}
    \end{subfigure}

    \vspace{0.2cm}

    \begin{subfigure}[t]{0.33\textwidth}
        \includegraphics[width=\linewidth]{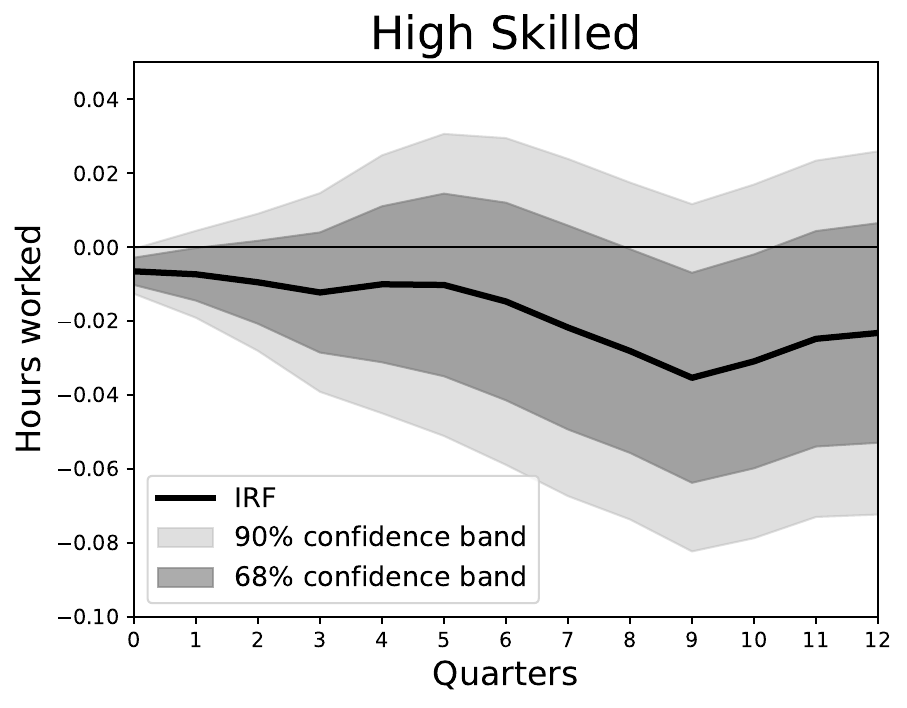}
    \end{subfigure}
    \hspace{0.1cm}
    \begin{subfigure}[t]{0.33\textwidth}
        \includegraphics[width=\linewidth]{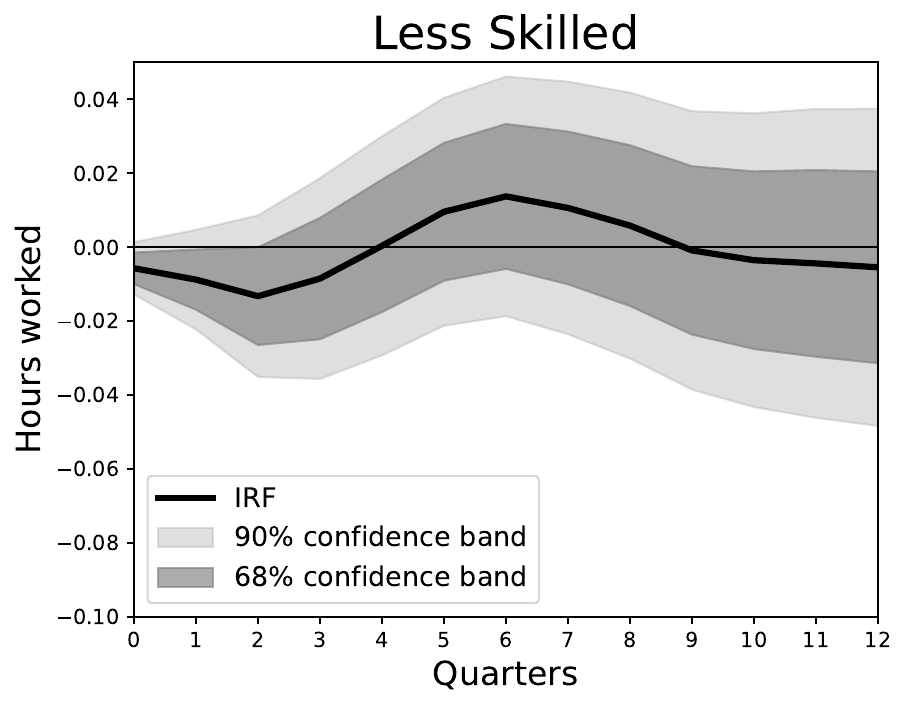}
    \end{subfigure}

    \caption{Skilled vs. unskilled: IRFs to a financial shock of weekly hours and wages}
    \vspace{0.15cm}
\footnotesize 
The figure shows the responses of wages (first line) and hours worked per week (second line) for high-skilled (left) and low-skilled (right) workers, after a one-standard-deviation positive financial shock. The solid line represents point estimates, while the shaded areas denote 90\% and 68\% confidence bands computed from panel HAC standard errors.
    \label{fig:skill_vs_low_skill_spec}
\end{figure}

\section{Robustness} \label{app_robustness}

\subsection{Not including uncertainity controls} \label{ineq_no_uncertainity_sec}

\begin{figure}[H]
    \centering
    \includegraphics[width=0.45\linewidth]{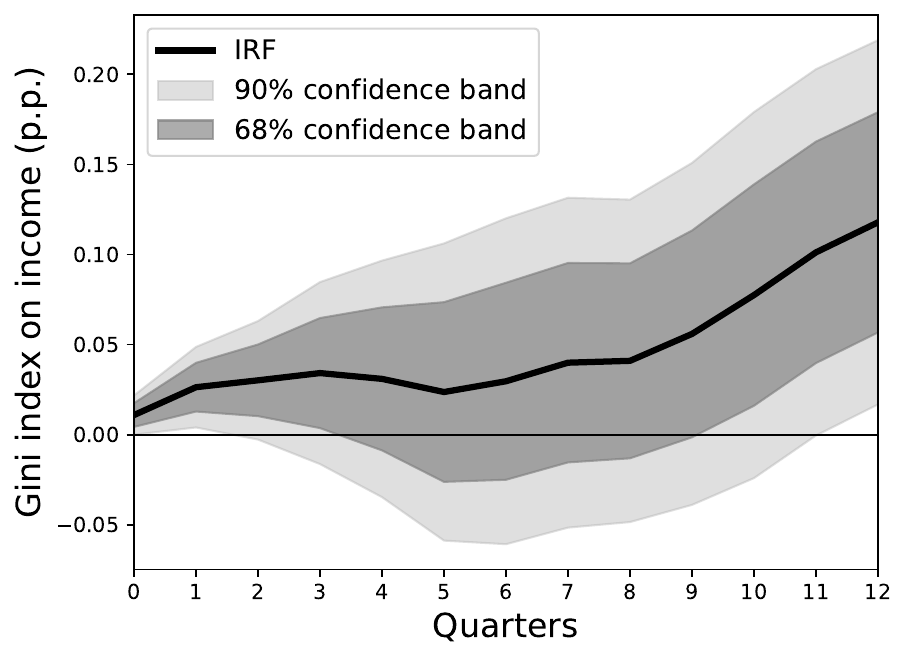}
    \caption{The effects of financial shocks on income inequality}
    \vspace{0.15cm}
\footnotesize 
The figure shows the response of income inequality after a one-standard-deviation positive financial shock occurs not considering the uncertainty controls. Full lines are point estimates; shaded areas indicate 90\% and 68\% confidence bands computed from panel HAC standard errors.
\label{fig:ineq_no_uncertainity}
\end{figure}

\subsection{Alternative identification strategy: recursive} \label{alternative_id}

\begin{figure}[H]
    \centering
    \includegraphics[width=0.45\linewidth]{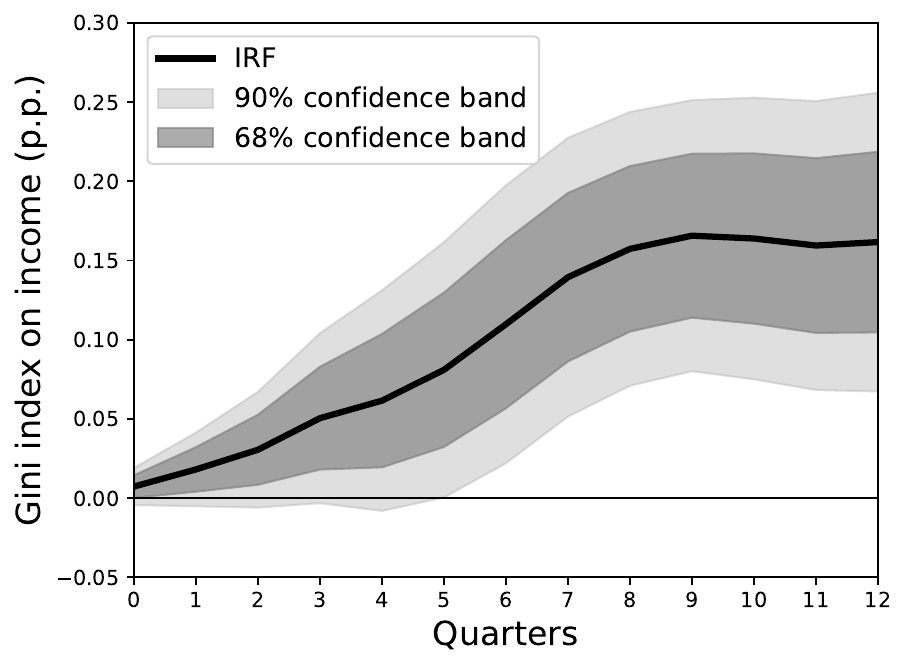}
    \caption{The effects of financial shocks on income inequality: recursive (Cholesky) identification}
    \vspace{0.15cm}
\footnotesize 
The figure shows the response of income inequality to a one-standard-deviation positive financial shock identified using a recursive ordering, with the share price variable placed last. The solid line represents point estimates, while the shaded areas denote the 90\% and 68\% confidence bands, computed using panel HAC standard errors.
\label{fig:chol}
\end{figure}

\subsection{Alternative inequality measures} \label{alternative_ineq_measures}

\begin{figure}[H]
    \centering
    \includegraphics[width=0.45\linewidth]{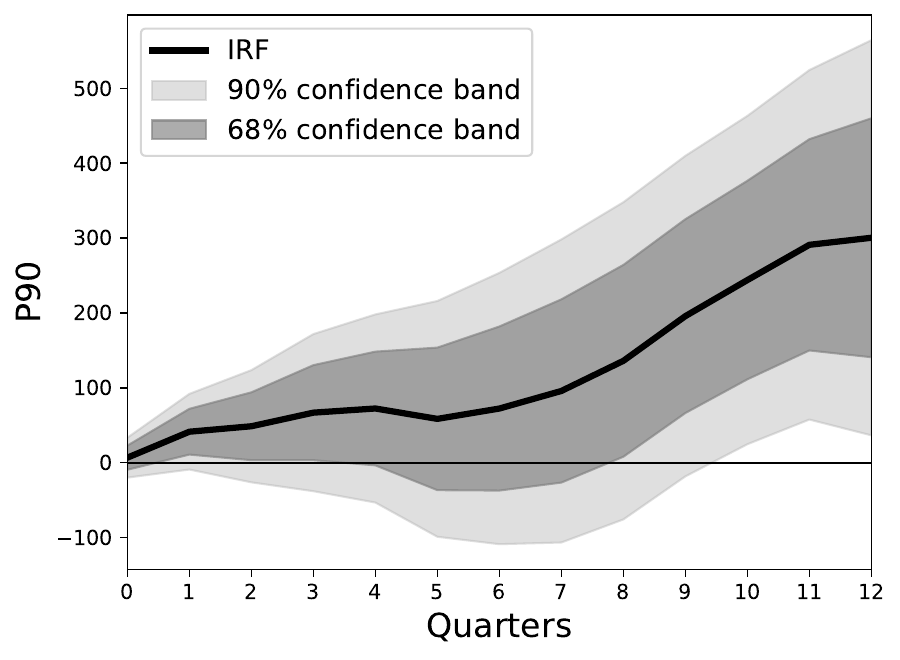}
    \caption{The effects of financial shocks on the 90th percentile}
    \vspace{0.15cm}
\footnotesize 
The figure shows the response of the 90th percentile to a one-standard-deviation positive financial shock. The solid line represents point estimates, while the shaded areas denote 90\% and 68\% confidence bands computed from panel HAC standard errors.
\label{fig:p90}
\end{figure}

\begin{figure}[H]
    \centering
    \includegraphics[width=0.45\linewidth]{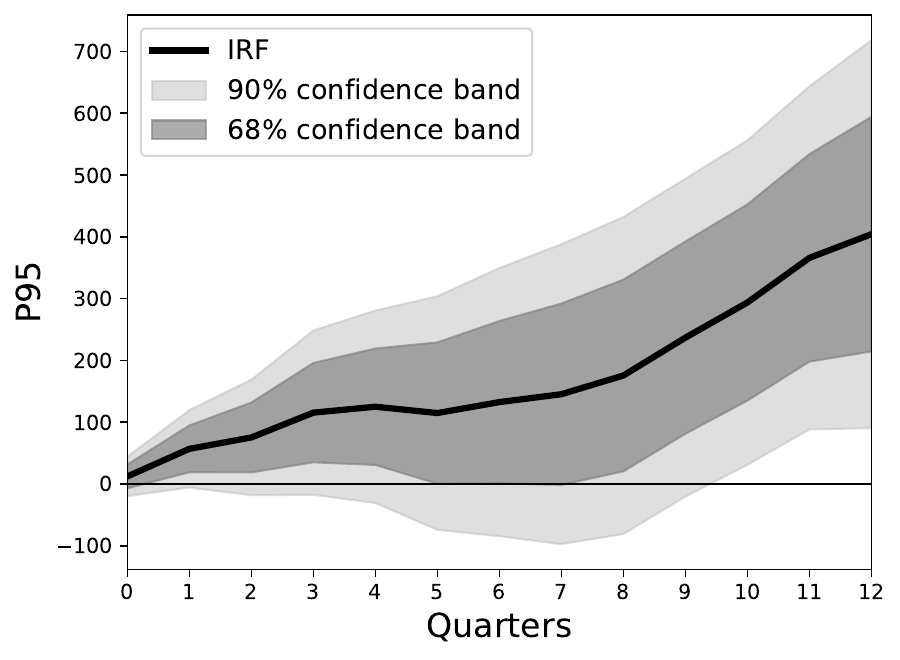}
    \caption{The effects of financial shocks on the 95th percentile}
    \vspace{0.15cm}
\footnotesize 
The figure shows the response of the 95th percentile to a one-standard-deviation positive financial shock. The solid line represents point estimates, while the shaded areas denote 90\% and 68\% confidence bands computed from panel HAC standard errors.
\label{fig:p95}
\end{figure}

\subsection{Subsample estimates: excluding the after covid-19 period}

\begin{figure}[H]
    \centering
    \includegraphics[width=0.45\linewidth]{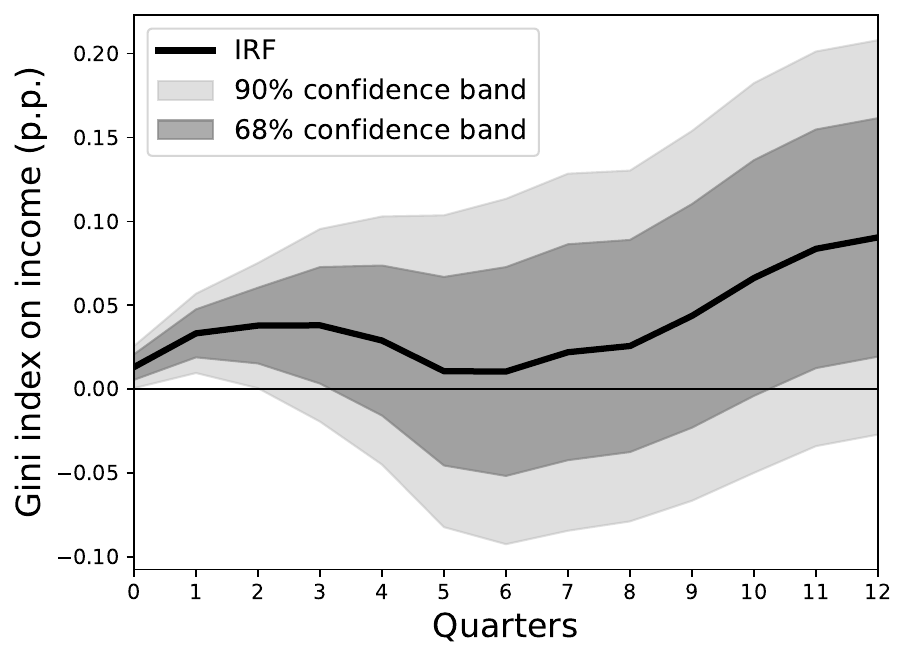}
    \caption{The effects of financial shocks on income inequality}
    \vspace{0.15cm}
\footnotesize 
The figure shows the response of income inequality to a one-standard-deviation positive financial shock. The solid line represents point estimates, while the shaded areas denote 90\% and 68\% confidence bands computed from panel HAC standard errors. The sample ends in 2019Q4.
    \label{fig:}
\end{figure}

\subsection{Subsample estimates: excluding extreme financial shocks} \label{excl_LV_NL}

\begin{figure}[H]
    \centering
    \includegraphics[width=0.45\linewidth]{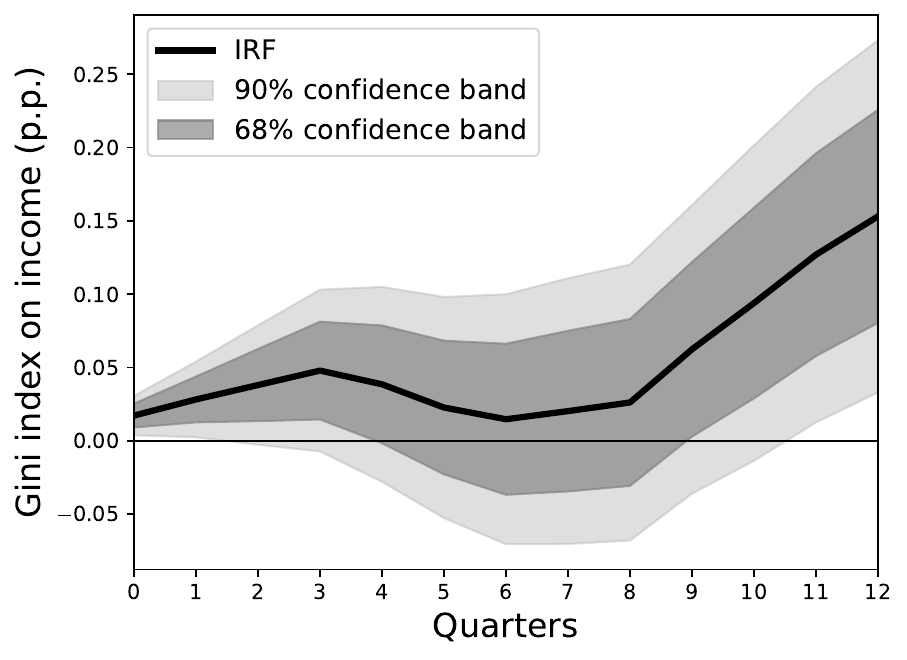}
    \caption{The effects of financial shocks on income inequality}
    \vspace{0.15cm}
\footnotesize  The figure shows the response of income inequality to a one-standard-deviation financial shock. The solid line represents point estimates, while the shaded areas denote 90\% and 68\% confidence bands computed from panel HAC standard errors. Latvia and the Netherlands are excluded from the sample.
    \label{fig:excl_LV_NL}
\end{figure}

\subsection{Credit shock} \label{credit}

\begin{table}[H]
\centering
\caption{Sign Restrictions}
\resizebox{0.98\textwidth}{!}{ 
\begin{tabular}{lcccccc}
\toprule
\textbf{}       & \textbf{Supply} & \textbf{Demand} & \textbf{Monetary} & \textbf{Investment} & \textbf{Financial} & \textbf{Credit}\\
\midrule
GDP             & +               & +               & +                 & +                   & +          & +        \\
Prices          & $-$              & +               & +                 & +                   & +          & +        \\
Interest rate   &                  & +               & $-$               & +                   & +          & +         \\
Investment/output &                & $-$             &                   & +                   & +          & +        \\
Stock prices    & +                &                 &                   & $-$                 & +          & +        \\
Credit/Stock prices           &                  &                 &                   &                     &  -         & +         \\
\bottomrule
\end{tabular}
}
\vspace{0.5cm}
\begin{minipage}{0.9\textwidth}
\small
\end{minipage}
\footnotesize
Signs represent imposed set-based restrictions at impact. A blank space indicates an unrestricted response.
\label{tab:restrictions_credit}
\end{table}

\begin{figure}[H]
    \centering
    \includegraphics[width=0.45\linewidth]{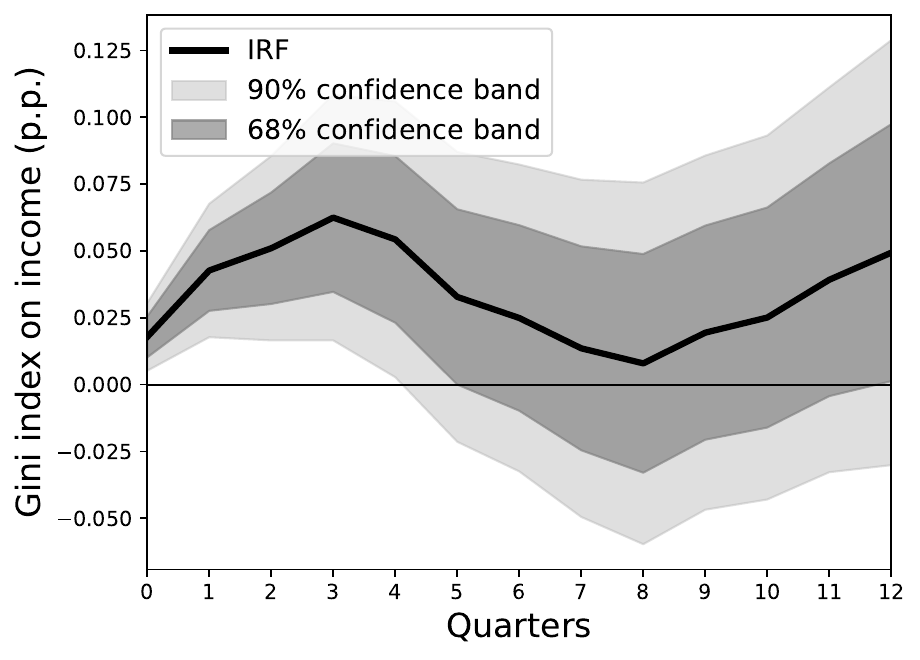}
    \caption{The effects of credit shocks on income inequality}
    \vspace{0.15cm}
\footnotesize  The figure shows the response of income inequality to a one-standard-deviation credit shock. The solid line represents point estimates, while the shaded areas denote 90\% and 68\% confidence bands computed from panel HAC standard errors.
    \label{fig:IRF_to_credit}
\end{figure}

\subsection{Alternative interpolation method}

\begin{figure}[H]
    \centering
    \includegraphics[width=0.45\linewidth]{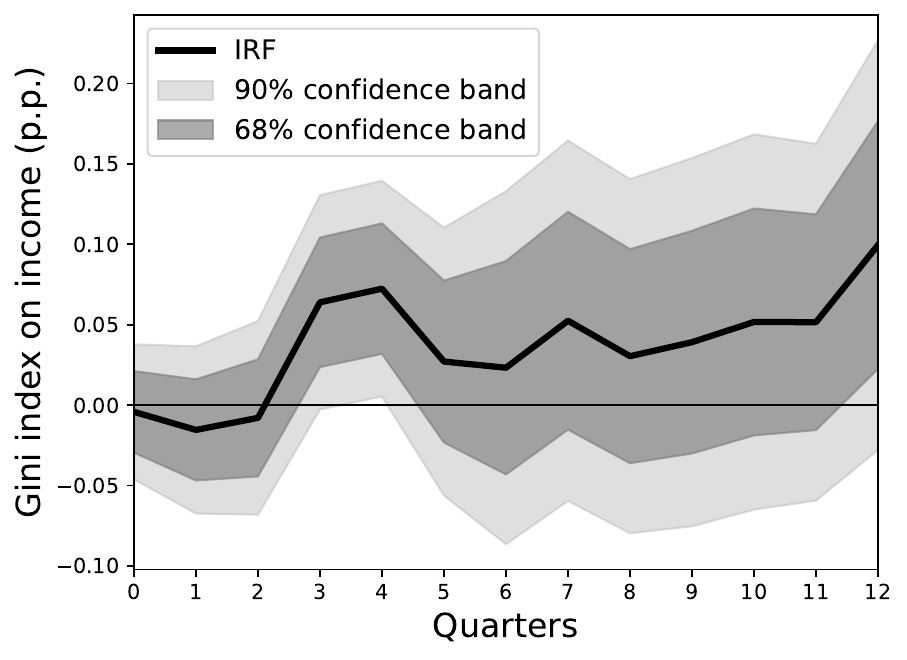}
    \caption{Income inequality response using flat interpolation}
\footnotesize 
The figure displays the response of income inequality to a one-standard-deviation positive financial shock, with the series converted to quarterly frequency using flat interpolation. The solid line represents the point estimate, while the shaded area indicates 68\% and 90\% confidence intervals computed from panel HAC standard errors.
    \label{fig:alt_trim_gini}
\end{figure}

\section{Nonlinear framework} \label{app_nonlin}
\begin{table}[H]
\centering
\caption{Sign Restrictions}
\resizebox{0.98\textwidth}{!}{ 
\begin{tabular}{lcccccc}
\toprule
\textbf{}       & \textbf{Supply} & \textbf{Demand} & \textbf{Monetary} & \textbf{Investment} & \textbf{Financial (positive)} & \textbf{Financial (negative)}\\
\midrule
GDP             & +               & +               & +                 & +                   & +          & -         \\
Prices          & $-$              & +               & +                 & +                   & +          & -         \\
Interest rate   &                  & +               & $-$               & +                   & +          & -         \\
Investment/output &                & $-$             &                   & +                   & +          & -         \\
Stock prices    & +                &                 &                   & $-$                 & +          & -         \\
Stock price volatility           &                  &                 &                   &                     &  +         & +         \\
\bottomrule
\end{tabular}
}
\vspace{0.5cm}
\begin{minipage}{0.9\textwidth}
\small
\end{minipage}
\footnotesize
Signs represent imposed set-based restrictions at impact. A blank space indicates an unrestricted response.
\label{tab:restrictions_id2}
\end{table}

\subsection{Alternative nonlinear specification}

\begin{figure}[H]
    \centering
    \includegraphics[width=0.45\linewidth]{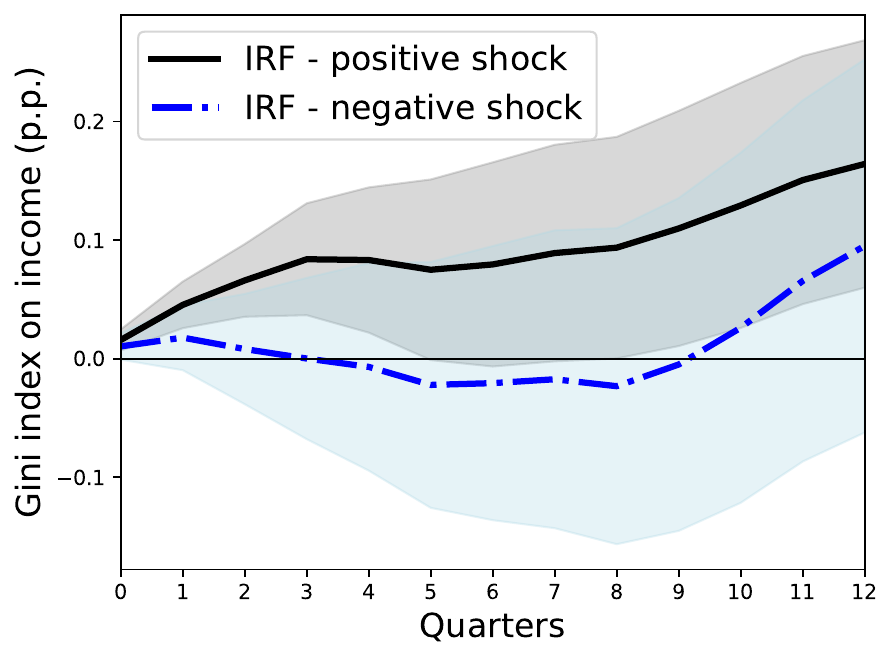}
    \caption{The asymmetric distributional effects of financial shocks}
    \vspace{0.15cm}
\footnotesize 
The figure shows the response of income inequality  to a one-standard-deviation positive (black) and negative (blue) financial shock. The solid line represents point estimates, while the shaded areas denote 68\% confidence bands computed from panel HAC standard errors. The nonlinearity here is modeled as in \textcite{tenreyro}.
    \label{fig:nonlinear_tenreyro}
\end{figure}

\section{Comparison of financial holdings in EU-SILC and HFCS data} \label{comparison}

To compare the distribution of financial income across the two datasets and assess potential underreporting in EU-SILC, we focus on the year 2017, which is available in both data sources, and on the same set of countries used in the primary analysis (excluding Lithuania, which is not considered in the HFCS).
\\The construction of the financial income variable was carried out as consistently as possible across datasets, following the definitions provided in the Methodological Guidelines and Description of EU-SILC Target Variables issued by the European Commission (2023 version). \\In the ECB dataset (HFCS), the relevant variable is “Gross Income from Financial Investment” (HG0410).

\begin{figure}[H]
    \centering
    \includegraphics[width=0.45\linewidth]{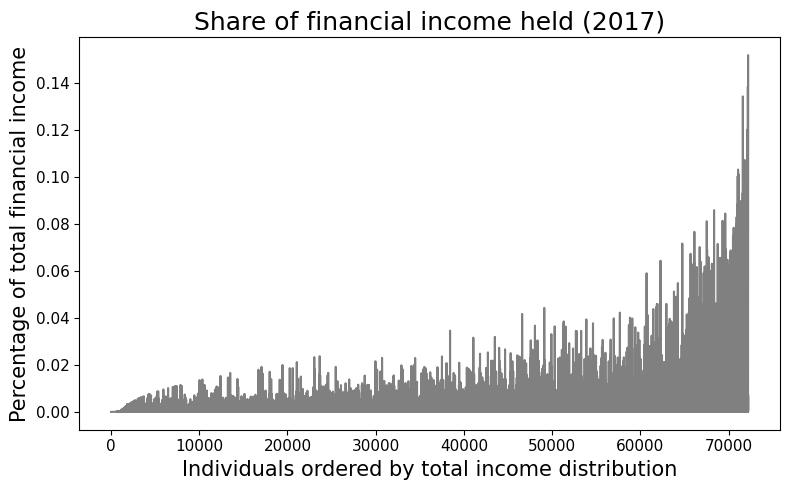}
    \caption{Financial income distribution in the Euro Area (2017) - EU-SILC}
    \vspace{0.15cm}
\footnotesize 
The figure shows the share of total financial income held by each individual, with individuals ranked according to their position in the overall income distribution. Top earners (those in the 99th percentile) are excluded from the graphical representation for visibility purposes. The underlying data are from the EU-SILC survey.
    \label{fig:fin_inc_distr_2017_eu_silc}
\end{figure}

\begin{figure}[H]
       \centering
    \includegraphics[width=0.45\linewidth]{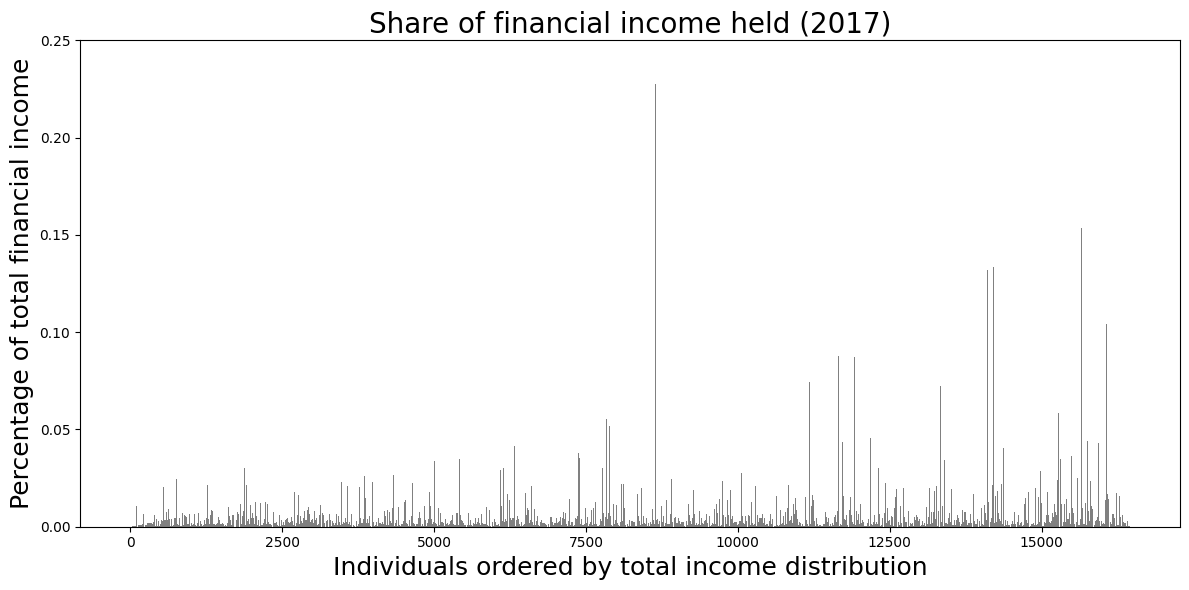}
    \caption{Financial income distribution in the Euro Area (2017) - HFCS}
    \vspace{0.15cm}
\footnotesize 
The figure shows the share of total financial income held by each individual, with individuals ranked according to their position in the overall income distribution. Top earners (those in the 99th percentile) are excluded from the graphical representation for visibility purposes. The underlying data are from the HFCS survey.
    \label{fig:fin_inc_distr_2017_eu_silc}
\end{figure}

\end{document}